\documentclass[a4paper,fleqn,usenatbib]{mnras}

\usepackage[T1]{fontenc}
\usepackage{ae,aecompl}

\usepackage{graphicx}	
\usepackage{amsmath}	
\usepackage{amssymb}	

\title[Chemical Abundances of Globular Clusters in NGC 5128]{Chemical Abundances of Globular Clusters in NGC 5128 (Centaurus A) 
\thanks{Based on observations made with ESO telescopes at the La Silla Paranal Observatory under programme ID 085.B-0107(A)}}

\author[S. Hernandez et al.]{
Svea Hernandez,$^{1}$\thanks{E-mail: s.hernandez@astro.ru.nl}
S{\o}ren Larsen,$^{1}$
Scott Trager,$^{2}$
Lex Kaper,$^{3}$
and Paul Groot$^{1}$
\\
$^{1}$Department of Astrophysics / IMAPP, Radboud University, PO Box 9010, 6500 GL Nijmegen, The Netherlands\\
$^{2}$Kapteyn Astronomical Institute, University of Groningen, Postbus 800, NL-9700 AV Groningen, the Netherlands\\
$^{3}$Astronomical Institute Anton Pannekoek, Universiteit van Amsterdam, Postbus 94249, 1090 GE Amsterdam, The Netherlands
}

\date{Accepted 2018 February 23. Received 2018 February 23; in original form 2018 January 22}

\pubyear{2017}

\begin{document}
\label{firstpage}
\pagerange{\pageref{firstpage}--\pageref{lastpage}}
\maketitle

\begin{abstract}
We perform a detailed abundance analysis on integrated-light spectra of 20 globular clusters (GCs) in the early-type galaxy NGC 5128 (Centaurus A). The GCs were observed with X-Shooter on the VLT. The cluster sample spans a metallicity range of $-1.92 < $ [Fe/H] $< -0.13$ dex. Using theoretical isochrones we compute synthetic integrated-light spectra and iterate the individual abundances until the best fit to the observations is obtained. We measured abundances of Mg, Ca, and Ti, and find a slightly higher enhancement in NGC 5128 GCs with metallicities [Fe/H] < $-$0.75 dex, of the order of $\sim$0.1 dex, than in the average values observed in the MW for GCs of the same metallicity. If this $\alpha$-enhancement in the metal-poor GCs in NGC 5128 is genuine, it could hint at a chemical enrichment history different than that experienced by the MW.  
We also measure Na abundances in 9 out of 20 GCs. We find evidence for intra-cluster abundance variations in 6 of these clusters where we see enhanced [Na/Fe] > $+$0.25 dex. We obtain the first abundance measurements of Cr, Mn, and Ni for a sample of the GC population in NGC 5128 and find consistency with the overall trends observed in the MW, with a slight enhancement ($<$0.1 dex) in the Fe-peak abundances measured in the NGC 5128.
\end{abstract}



\section{Introduction}
There is a bewildering variety of properties in the galaxies we currently observe in the universe. The differences between these objects range from morphology, luminosity, and colour to star formation histories, chemical composition and kinematics. In order to fully understand how galaxies in our universe evolve we need to study in detail galaxies of different Hubble types, not just our own.  \par
A very powerful tool to understand the evolution of galaxies is the study of their chemical abundances as a function of time. More specifically, one can obtain a detailed picture of the evolution of galaxies by looking closely at the abundance patterns of different elements observed in various stellar populations. Considering that the chemical composition of the gas reservoirs that form stars is preserved in their atmospheres,  one can extract an immense amount of information through the analysis of stars of different ages. Past studies have shown that the stellar abundance ratios can constrain initial mass functions (IMFs) and star formation rates  \citep[SFR, ][]{mc97,mat03}. \par
It is believed that type Ia supernovae (SN Ia) are the primary sources of Fe and Fe-peak elements in the Galaxy \citep{nom97}.  Additionally, it has been established that type II supernovae (SN II) are instead responsible for the creation of most of the $\alpha$-elements \citep[O, Mg, Si, S, Ca, and Ti, ][]{woo95}. \par 
One of the most widely used diagnostics of IMF and SFR is the abundance ratio of $\alpha$-elements to Fe, [$\alpha$/Fe]. A top-heavy IMF could lead to enhanced [$\alpha$/Fe] ratios along with high average metallicities \citep{mat90}. However, similar enhancements in the [$\alpha$/Fe] ratios can also be the result of a rapid burst of star formation enriching the gas to high metallicities. In the case of our own Milky Way (MW) the enhancement of $\alpha$-elements in old stellar populations points at a starburst system \citep{wor98, mat03}. \par
Given our proximity to stars in the MW, high-resolution spectroscopy has allowed for abundance studies that provide an  incredibly detailed picture of the nucleosynthetic history of our Galaxy. Individual stellar abundances within the MW  have assisted in identifying the location and substructures of different populations \citep[e.g. ][]{ven04, pri05, red06}. Even though a limited number of extragalactic abundances of stars in nearby systems, such as the Magellanic Clouds and dwarf galaxies, are also available \citep[e.g. ][]{wol73, ven99, she98, tol03}, to better understand the episodes of star formation in different galaxy types and masses, detailed abundances far beyond the MW and its neighbours in the Local Group are needed. \par
Most of the abundance work beyond the Local Group has been limited by the difficulty in measuring reliable abundances. The majority of the extragalactic (outside of the Local Group) metallicity and abundance measurements come from studies of H II regions in star-forming galaxies \citep{sea71,lee04, sta05}. A known issue with such methods is apparent once we compare the metallicities inferred from different diagnostics. Studies have noticed differences in the inferred metallicities as high as $\sim$0.7 dex \citep{ken03,bre08, kew08, lop12}. Additionally, H II region abundances mainly probe the present-day gas composition, and one can not access any information on the past evolution of the host galaxy. \par
For galaxies other than star-forming, different methods are used to study their composition, and star formation histories. For early-type galaxies several extensive studies have been conducted using absorption line indices as their main resource. \citet{tho05} studied the stellar properties of 124 early-type galaxies deriving metallicities, [$\alpha$/Fe] ratios, and ages using absorption line indices as well as stellar population models. They find that all three, age, metallicity and [$\alpha$/Fe] abundance ratio, correlate with the mass of the galaxy. These results, especially those of the [$\alpha$/Fe]-mass relation had been anticipated before the work of \citet{tho05} by \citet{wor92}, \citet{fis95}, and \citet{kun00}, amongst others, and verified by \citet{tra00}, \citet{pro02}, and \citet{tho02} through stellar population models. Overall, this relation between [$\alpha$/Fe] and mass observed in earlier studies is consistent with a scenario where higher effective yields from SNe II are expected and observed in more massive galaxies ($>$ 10$^{11}$ M$_{\odot}$). \par
One of the main challenges in obtaining detailed abundances of individual stars in galaxies outside of the Local Group is the fact that at larger distances stars become too faint for this type of analysis. In order to overcome this obstacle, studies are now focusing on star clusters, where one of the assumptions is that the individual populations of stars consist of objects of the same age, and are chemically homogenous. The study of integrated-light observations has allowed for a significant advancement in the field of chemical evolution of distant galaxies. Given that the integrated-light spectra of most star clusters are broadened by several km s$^{-1}$, one can observe star clusters at much higher resolution ($R=$20,000-30,000) than galaxies, allowing for the detection of weak lines (15 m\r{A}). Studies of star clusters, mainly globular clusters (GC), have shown that these objects trace the properties of the different field star populations in their host-galaxies \citep{col13, sak15}. More specifically, detailed properties such as metallicity, age and abundances are powerful tools for constraining theories of GC and galaxy formation \citep{bro06}. \par
One of the major highlights in the studies of extragalactic GCs is the presence of subpopulations with two main components, metal-poor and metal-rich. Until the early 2000s, most of the spectroscopic analysis exploring extragalactic GCs utilised Lick/IDS indices \citep{bur84,wor94,tra98}. This technique was originally developed to study the absorption features in low resolution observations at wavelengths of $\sim$4000-6400 \r{A} of early-type galaxies to measure the properties of their stellar populations. More optimal methods based on the index system were later developed to be more applicable to GC studies. The bulk of extragalactic spectroscopic measurements of metallicities, ages and [$\alpha$/Fe] ratios have been inferred mainly by measuring Lick indices. One of the most extensive and systematic studies of extragalactic GC systems was published by \citet{str05} and spanned a broad range of galaxies from dwarfs to ellipticals. This work showed that both GC subpopulations, metal-poor and metal-rich, have mean ages similar to those seen in the MW GC system with the implication that the bulk of star formation in spheroids occurred at early ages ($z$> 2). \par
The general expectation regarding the metal-poor and metal-rich populations of GCs in external galaxies is that the former would show supersolar [$\alpha$/Fe] ratios given that their formation took place in the early universe when a substantial metal enrichment was yet to happen. Several studies of extragalactic metal-poor GCs appear to have [$\alpha$/Fe] $\lesssim$ 0 \citep{ols04, pie05}, while others show enhanced [$\alpha$/Fe]. On the other hand, a study of GCs in early-type galaxies by \citet{puz05}, using Lick indices to measure ages, metallicities and [$\alpha$/Fe] abundances leads them to conclude that [$\alpha$/Fe] ratios are on average super-solar with a mean value of $+$0.47 $\pm$ 0.06 dex which points at short star formation timescales ($\sim$ 1 Gyr). They also conclude that the progenitor cloud forming these GCs in early-type galaxies would need to have been predominantly enriched by yields from SNe II. The discrepancy between results obtained from different studies of early-type galaxies could be caused by the uncertainties in the simple stellar population (SSP) models used in combination with the wide Lick index bandpasses, making the abundance ratio measurements highly uncertain \citep{bro06}. \par
\citet{mc08} developed a technique to study individual abundances of GCs through the analysis of their integrated-light. Their method requires high-resolution observations ($R$=30,000) and combines information of the the Hertzsprung-Russell diagrams (HRD) of the GC in question, stellar atmospheric models and synthetic spectra. This technique has been tested and applied to GCs in the MW and the Large Magellanic Cloud \citep[LMC, ][]{mc08,cam09,col11}, and at relatively larger distances \citep[M31 at $\sim$ 780 kpc, ][]{col09,col14,col16}.\par

Similar to the concept of \citet{mc08}, \citet[][, hereafter L12]{lar12} created a high-resolution integrated-light technique to measure abundances of star clusters. In contrast to  the technique of \citet{mc08}, the method of L12 mainly relies on spectral synthesis and full spectral fitting. Abundance measurements are obtained by fitting relatively broad wavelength ranges containing multiple lines of the element in question. The method of L12 has been used to measure chemical abundances of old stellar populations in the MW, and several dwarf spheroidal galaxies \citep{lar12,lar14, lar17}. \par

Although the L12 technique was developed using high-dispersion spectroscopic observations, we extended the L12 method to intermediate-resolution spectra ($R < 10,000$) in \citet{her17} and \citet{her18}. In these studies we measured detailed abundances of two young massive clusters (YMC) and overall metallicities of eight YMCs in galaxies $\sim$5 Mpc away. With techniques like L12 where one is able to study stellar populations in a broad range of ages, from old GCs to young populations (YMCs), and with a slightly less limited spectral resolution range, we can now investigate the chemical evolution of galaxies through a much larger window in space and time than in past studies.\par

Being the nearest giant early-type galaxy to the MW, NGC 5128 (Centaurus A) is an ideal target to measure detailed abundances from integrated-light spectra of its globular clusters. In spite of being the nearest giant early-type galaxy, NGC 5128 is still located at a distance of 3.8 Mpc from our Galaxy \citep{har10}. \citet{col13} initiated a study of the chemical composition of the GC system of NGC 5128 measuring Fe and Ca abundances of 10 different clusters. In their work \citet{col13} measured Fe abundances with ranges of $-1.6 <$ [Fe/H] $< -0.2$ dex. A noticeable result from this work was the enhanced [Ca/Fe] ratio for metallicities of [Fe/H] $<-0.4$ dex, which appeared higher than the average values for the GCs in the MW and M31 of the same metallicities. These results implied a different star-formation history for NGC 5128 compared to those from the MW and M31. \par 

We perform a detailed abundance analysis on a sample of 20 star clusters distributed throughout NGC 5128. Using the L12 method, along with intermediate-resolution observations we measure abundances of light- (Na) and $\alpha$ elements (Mg, Ca and Ti), as well as Fe-peak (Cr, Mn, and Ni). This paper is structured as follows. In Section \ref{sec:obs} we describe the science observations and data reduction. In Sections \ref{ana} and \ref{results} we present the analysis method and results, respectively. Section \ref{discuss} is used to discuss our findings, and we list our conclusions in Section \ref{con}.


\section{Observations and Data Reduction}\label{sec:obs}
In this work we exploit the data taken as part of the VLT programme 085.B-0107(A) observing 25 bright star clusters in NGC 5128. The exposures are taken with the X-Shooter single target spectrograph \citep{ver11}. The instrument has a broad wavelength coverage, providing data between 3000-24800 \r{A}. X-Shooter's extensive wavelength coverage is possible through its three spectroscopic arms. Each of these arms, UV-Blue (UVB), Visible (VIS), and Near-IR (NIR), has a full set of optimised optics and detectors. The resolution of the instrument is mainly controlled by the slit width, with resolutions ranging from R=3000 to 17000. Programme 085.B-0107(A) was executed in April 2010 using slit widths 0.8\arcsec (R$\sim$6200), 0.7\arcsec (R$\sim$11000),  and 0.6\arcsec (R$\sim$6200) for UVB, VIS, and NIR arm, respectively. The data was taken using the standard nodding mode following an ABBA pattern. \par
\begin{figure}
   \resizebox{\hsize}{!}
            {\includegraphics[width=11.2cm]{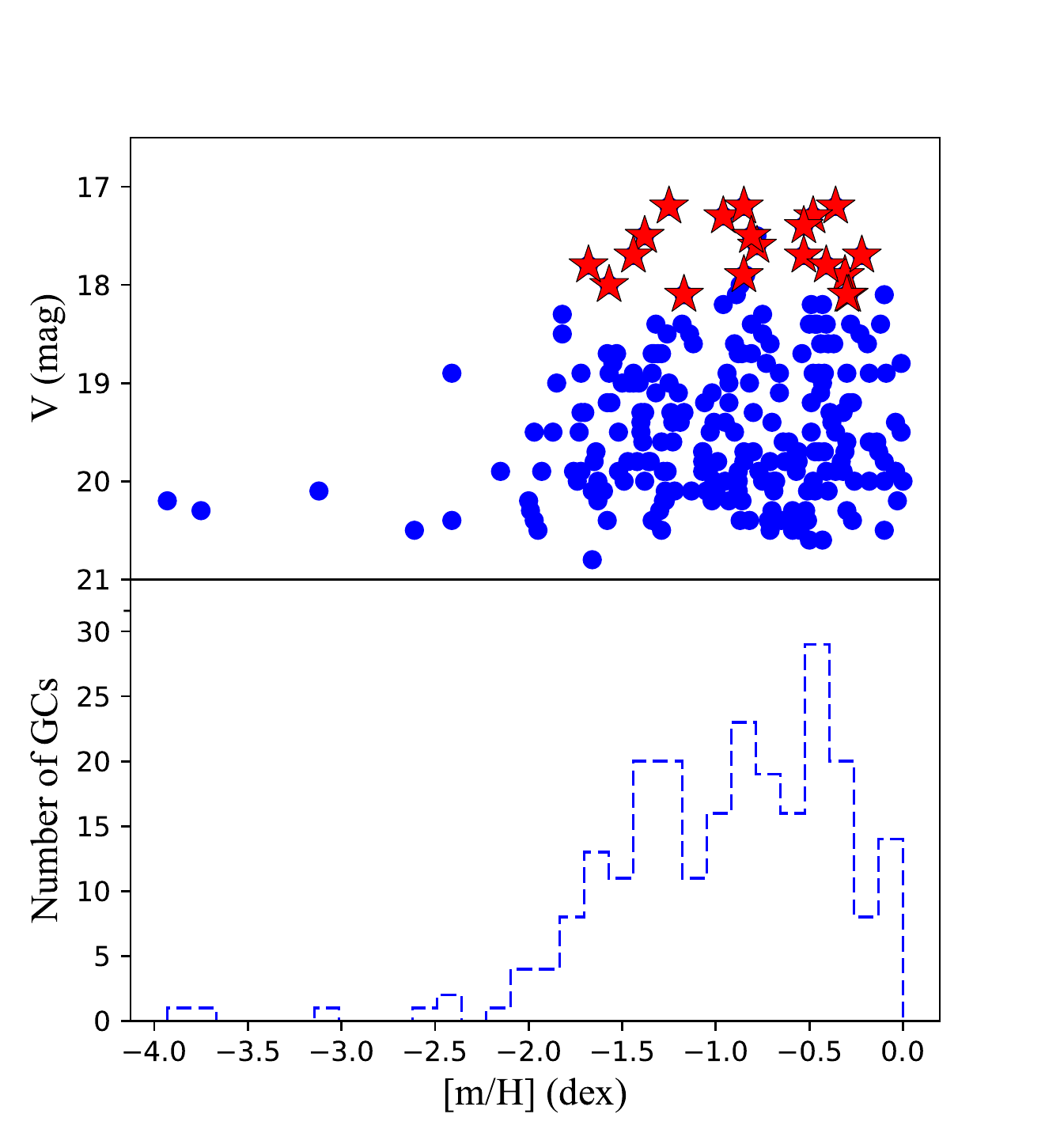}}
      \caption{Top panel: V magnitudes as a function of spectroscopic metallicities as measured by \citet{bea08}. Blue circles and red stars show the full GC study by \citet{bea08} and the sample studied in this work, respectively. Bottom panel: Metallicity distribution (bin widths of $\sim$ 0.13 dex) for the GC system in NGC 5128.}
         \label{fig:z_mag}
   \end{figure}
\begin{figure}
   \resizebox{\hsize}{!}
            {\includegraphics[width=11.2cm]{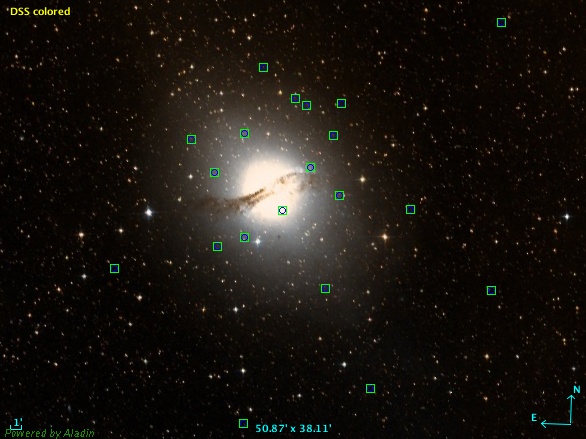}}
      \caption{Colour composite image from the DSS2 optical HEALpix survey, including $\sim$0.6$\micron$ and $\sim$0.4$\micron$  bands. Green squares locate the clusters studied as part of this work. }
         \label{Fig:galaxy}
   \end{figure}

Targets with magnitudes brighter than $V\sim19$ were selected from the spectroscopically confirmed GC sample by \citet{bea08}. This means that our cluster sample mainly covers the brightest component of the GC system in NGC 5128. In Figure \ref{fig:z_mag} we show in blue circles the full GC sample of \citet{bea08}, plotting the V magnitudes as a function of metallicity. We indicate with red stars the cluster sample analysed in this work. In Table ~\ref{table:obs} we list the target names, coordinates, exposure times and signal-to-noise (S/N) ratios for the different arms. The S/N values are calculated using wavelength windows of 4550-4750 \r{A} for the UVB, 7350-7500 \r{A} for the VIS data and 10400-10600 \r{A} for the NIR data. We point out that given the low S/N in the NIR data, the work done in this paper focuses on the X-Shooter observations taken with the UVB and VIS arm only. In Figure ~\ref{Fig:galaxy} we mark the star clusters analysed in this paper.\par

Using the public release of the X-Shooter pipeline (v2.5.2) along with the ESO Recipe Execution Tool (EsoRex) v3.11.1 we perform the basic data reduction steps, including bias and dark corrections, flat-fielding, sky subtraction and wavelength calibration. We take the two-dimensional calibrated exposures and extract the science spectrum using IDL routines by \citet{che14}. The code was developed based on optimal extraction principles described in \citet{hor86}. Once the 1-D spectrum is extracted, the code combines the individual orders using the variance-weighted average for overlapping spectral regions.  
The 1-D spectra are flux calibrated using the spectrophotometric standard Feige 110 observed close in time to the science frames. We use the pipeline routine xsh\_respon\_slit\_offset to create response curves for the individual exposures. The  response curves use the same bias and master flat frames as those applied to the corresponding science observations, and are corrected for exposure time and atmospheric extinction. It is important to apply the same flat field frames for both the science and the response curves in order to remove any contemporaneous flat-field features. After the flux calibration, we visually inspect the 1-D spectra and find good flux agreement between the UVB and VIS arm. We note that in most of the targets observed in this program we see strong dichroic features at wavelengths $<$ 5700 \r{A} (in the VIS arm). These dichroic features tend to appear in the extracted 1-D spectra of the X-Shooter VIS data, with their exact wavelength position varying from exposure to exposure making it rather difficult to remove entirely \citep{che14}. Although our abundance analysis does not require accurate flux calibration, these dichroic features can artificially modify  not only the continuum, but also the depth of the intrinsic spectral lines. Given the nature of the dichroic features, we exclude most of the VIS wavelengths $<$ 5700 \r{A} in our analysis. \par

Telluric absorption bands strongly affect our VIS exposures. Observations with wavelengths between 5800-6100 \r{A}, 6800-7400 \r{A}, 7550-7750 \r{A}, 7800-8500 \r{A}, and 8850-10 000 \r{A}, are mainly contaminated by these absorption bands. We correct for this contamination from the Earth's atmosphere using the routines and telluric library developed by the X-Shooter Spectral Library (XSL) team. \citet{che14} created a correction method based on Principal Component Analysis (PCA) which depends on their carefully designed telluric library. We apply this correction which reconstructs and removes the strongest telluric absorption as needed. Although we correct our observations for telluric contamination, in our analysis we aim to avoid these regions. We mainly focus  our abundance analysis in the UVB wavelengths 4400-5200 \r{A}, and VIS wavelength windows 5670-5700 \r{A}, 6100-6800 \r{A}, 7400-7550 \r{A}, and 8500-8850 \r{A}.

\begin{table*}
\caption{X-Shooter Observations}
\label{table:obs}
\centering 
\begin{tabular}{cccccccccc}
\hline \hline
Cluster & RA & Dec &  \multicolumn{3}{c}{$t_{\rm exp}$ (s)} & \multicolumn{3}{c}{S/N (pix$^{-1}$)}   \\
 & (J2000) & (J2000) & UVB  & VIS  & NIR & UVB & VIS & NIR \\
\hline\\
AAT117062 & 201.378408 & -42.83718 & 1800.0 & 1780.0 & 1800.0 &  29.1 & 34.0 & 13.5  \\
HGHH-04 & 201.256648 & -43.15735 & 1800.0 & 1780.0 & 1800.0 & 12.3 & 18.1 & 7.2 \\
HGHH-06 & 201.342347 & -43.04529 & 1800.0 & 1780.0 &  1800.0 & 14.6 & 19.2 & 8.6 \\
HGHH-07 & 201.521448 & -42.94223 & 1800.0 & 1780.0 &  1800.0 & 25.9 & 29.4 & 12.0 \\
HGHH-11 & 201.228781 & -43.0223 & 1800.0 & 1780.0 &  1800.0 & 16.9 & 22.4  & 10.7\\
HGHH-17 & 201.416499 & -42.93268 & 1800.0 & 1780.0 & 1800.0 &  24.3 & 28.1 & 11.5 \\
HGHH-21 & 201.470672 & -43.09595 & 1800.0 & 1780.0 & 1800.0 & 18.5 & 22.7 & 9.0\\
HGHH-23 & 201.476303 & -42.99024 & 1800.0 & 1780.0 & 1800.0 &  28.6 & 39.6 & 24.0 \\
HGHH-29 & 201.167061 & -43.30198 & 1800.0 & 1780.0 & 1800.0 & 12.1 & 20.0  & 10.6\\
HGHH-34 & 201.41993 & -43.3532 & 1800.0 & 1780.0 &  1800.0 & 13.2 & 16.7  & 7.9 \\
HGHH-40 & 200.927402 & -43.16027 & 1800.0 & 1780.0 & 1800.0 &  8.8 & 11.1 & 3.4 \\
HH-080 & 200.909807 & -42.77231 & 1800.0 & 1780.0 & 1800.0 & 14.9 & 17.1  & 5.6\\
HH-096 & 201.088942 & -43.04287 & 1800.0 & 1780.0 &  1800.0 & 18.5 & 21.5  & 8.2\\
HHH86-30 & 201.225916 & -42.88951 & 1800.0 & 1780.0 & 1800.0 & 31.2 & 37.9 & 19.8 \\
HHH86-39 & 201.674945 & -43.12846 & 1800.0 & 1780.0 & 1800.0 & 32.0 & 34.9 & 13.5 \\
K-029 & 201.287304 & -42.98286 & 1800.0 & 1780.0 &  1800.0 & 17.6 & 24.1 & 11.5\\
K-034 & 201.293745 & -42.89216 & 1800.0 & 1780.0 &  1800.0 & 20.4 & 26.2  & 12.5 \\
K-163 & 201.416752 & -43.08328 & 1800.0 & 1780.0 &  1800.0 & 27.2 & 31.6  & 14.2\\
VHH81-03 & 201.241579 & -42.93562 & 1800.0 & 1780.0 & 1800.0 &  18.8 & 24.7 & 13.9 \\
VHH81-05 & 201.316943 & -42.88218 & 1800.0 & 1780.0 &  1800.0 & 20.6 & 22.0 & 7.5 \\
\hline
\end{tabular}
\end{table*}


\section{Abundance Analysis}\label{ana}
In this work we make use of the integrated-light analysis tool developed and tested by L12. The L12 technique was originally created for high-dispersion observations of GCs, however in \citet{her17} we showed that this method can also be used for detailed abundance analysis of intermediate-resolution observations. The original method is described in detail in the work of \citet{lar12} and \citet{lar14}. The main idea involves an iterative process where the chemical abundances are determined by fitting synthetic modelled spectra to the integrated-light observations, varying the abundances on each iteration. To accurately model the observations one needs to account for every evolutionary stage present in the star cluster in question. We compute a single model atmosphere, which is then used to generate a  high-resolution ($R\sim 500,000$) synthetic spectrum for each stellar type. The individual model spectra are then combined onto a single synthetic integrated-light spectrum. We note that when co-adding the spectra we use appropriate weights accounting for the number of stars of a specific type based on a Salpeter IMF (see Section \ref{theo} for a detailed description). The synthetic integrated-light spectra are smoothed to match the resolution of the X-Shooter observations, and then compared to the science data.\par
In the original work of L12 the software made use of \textsc{atlas}9 models along with \textsc{synthe} codes to compute the model atmospheres and synthetic spectra \citep{kur70, kur79, kur81}. \textsc{atlas}9 is a one-dimensional (1D) plane-parallel local thermodynamic equilibrium (LTE) atmospheric modelling software by Robert Kurucz. However, these plane-parallel models are less ideal for cooler stars. In our analysis in addition to using \textsc{atlas}9 and \textsc{synthe} codes, we also use \textsc{marcs} atmospheric models \citep{gus08} paired with \textsc{turbospectrum} to create the synthetic spectra \citep{ple12}. In contrast to \textsc{atlas}9, \textsc{marcs} are 1D spherical LTE models. We download a grid of precomputed \textsc{marcs} models from their official website\footnote{http://marcs.astro.uu.se}, and allow the code to select the closest model to match the different stellar types. We use \textsc{atlas}9/\textsc{synthe} software for stars with $T_{\rm eff}$ > 5000K and \textsc{marcs/turbospectrum} for stars with $T_{\rm eff}$ < 5000K, similar to what was done in \citet{her18}. Our analysis is based on the Solar composition from \citet{gre98}.  \par
The abundance analysis done in this paper is entirely based on LTE modelling. We note that we do not apply any non-LTE (NLTE) corrections as these are rather complex for integrated-light work given their dependance on the stellar properties of individual stars. In general NLTE corrections for some $\alpha$-elements are predicted to range between $-0.4$ and $-0.1$ dex \citep{ber15}.

\subsection{Theoretical+Empirical stellar parameters}\label{theo}
The distance to NGC 5128 makes obtaining colour-magnitude diagrams (CMDs) a complicated task. Given that CMDs for our star-cluster sample are not available, we instead rely on theoretical isochrones for the analysis. The selection of the atmospheric models is based on the theoretical $\alpha$-enhanced isochrones by the Dartmouth group \citep{dot07} which account for main sequence (MS) and red giant branch (RGB) stars. We note that these isochrones exclude both, the horizontal branch (HB) and asymptotic giant brach (AGB) stages. For this reason we opt for combining the theoretical isochrones with empirical HB and AGB observations from the survey of Galactic GCs using the Advanced Camera for Surveys (ACS) on board the Hubble Space Telescope by \citet{sar07}. \par
For the initial selection of isochrones and appropriate empirical HB and AGB data we adopt the ages and metallicities from \citet{bea08} listed in Table \ref{table:age_fe}. We extract the stellar parameters from the individual isochrones assuming an IMF following a power law with a \citet{sal55} exponent, $\alpha$=$-$2.35. We adopt the luminosity limit of L12 where we only include stars brighter than $M_{\rm V}$ = +9. L12 points out that including stars fainter than this limit modifies the overall metallicities by $<$ 0.1 dex. \par
We also choose a CMD with similar age and metallicity as the ones listed in Table \ref{table:age_fe} and extract the photometry for stars in the HB and the AGB. For each of the GCs in NGC 5128 we list the Galactic GC used to model the HB and AGB populations in the last column of Table \ref{table:age_fe}. The photometry is corrected for foreground extinction using the colour excess values, $E(B-V)$, from the latest edition of the \citet{har96} catalogue (version 2010). Most of the distance estimates used in the photometry come from the work of the 2010 edition of \citet{har96} and \citet{van06}. For the HB and AGB stars we derive values for the $T_{\rm eff}$ and bolometric corrections from the $V-I$ colours using the colour-$T_{\rm eff}$ transformation based on models by Kurucz \citep{cas97}. Additionally, the surface gravities, log $g$, are inferred using the relation\par
\begin{equation}
\log g\; =\; \log\: g_{\odot}\; +\; \log\: \bigg[ \bigg(\frac{T_{\rm eff}}{T_{\rm eff \odot}}\bigg)^{4} \bigg(\frac{M}{M_{\odot}}\bigg)  \bigg(\frac{L_{\rm bol}}{L_{\rm bol \odot}}\bigg)^{-1}\bigg]
\end{equation}
When combining the isochrone-based CMDs with the empirical HB and AGB data, similar to the approach of \citet{lar17}, we estimate the weights of the empirical data by matching the number of RGB stars in the range of $1< M_{V}<2$ in the theoretical CMD to those present in the empirical CMD.\par
We also account for the microturbulent velocity component, $v_{t}$, assigning different values depending on their $\log g$. For stars with $\log g$ > 4.5 we adopt a value of $v_{t}$=0.5 km s$^{-1}$ \citep{tak02}. Following the reference points of L12 and \citet{lar17}, for 4.0 < $\log g$ < 4.5 we assume $v_{t}$=1.0 km s$^{-1}$ and for $\log g$ < 1.0,  $v_{t}$=2.0 km s$^{-1}$. For values 1.0 < $\log g$ < 4.0 we assign microturbulent velocities based on a linear interpolation of  $v_{t}$($\log g$). Lastly, for HB stars we assume $v_{t}$=1.8 km s$^{-1}$ \citep{pil96}. \par

\subsection{Smoothing parameter}
After the synthetic integrated-light spectrum is produced we degrade the resolution of the model spectrum from $R=500,000$ to $R\sim6,200 - 11,000$ to match the X-Shooter observations. In our analysis we are able to fit for the best Gaussian dispersion, $\sigma_{\rm sm}$, which should account for the instrumental resolution ($\sigma_{\rm inst}$) and the cluster velocity dispersion ($\sigma_{\rm 1D}$).\par
As a first step in this analysis, we fit for the radial velocities ($v_{\rm rv}$) and the best $\sigma_{\rm sm}$ along with the overall metallicity, [Z], processing 200 \r{A} of data at a time. Given that X-Shooter collects data through a multiple-arm system detailed in Section \ref{sec:obs}, we fit separate $\sigma_{\rm sm}$ for each arm. We scan the UVB wavelength range between 4000 \r{A} to 5200 \r{A}, and for the VIS arm we use the spectroscopic observations covering wavelengths between 6100 \r{A} to 8850 \r{A} excluding areas affected by strong telluric absorption. \par
Using slit widths of 0.5\arcsec\: for the UVB arm and 0.7\arcsec\: for the VIS arm \citet{che14} find that the instrumental resolution in the UVB arm varies between $R= 9,584-7,033$. In contrast to their findings regarding the UVB observations, the VIS arm showed a constant resolution throughout the wavelength coverage with an average value of $R=10,986$, very similar to the X-Shooter stated resolution of $R=11,000$\footnote{https://www.eso.org/sci/facilities/paranal/instruments/\\xshooter/inst.html}. Following the results of \citet{che14}, we assume that the resolving power follows a Gaussian full width half maximum (FWHM), and a constant resolution in the VIS arm with a value of $R=11,000$, corresponding to an instrumental velocity dispersion of $\sigma_{\rm inst}=11.58$ km s$^{-1}$. \par
We estimate the line-of-sight velocity dispersions by subtracting $\sigma_{\rm inst}$ in quadrature from the averaged $\sigma_{\rm sm}$ for the VIS data. We note that the work presented here assumes an instrumental resolution set by the configuration alone, i.e. slit width.  In Table \ref{table:derived_vel} we list our inferred radial velocity and the line-of-sight velocity dispersions for each of the clusters in the sample. \par
\begin{figure*}
            {\includegraphics[scale=0.60]{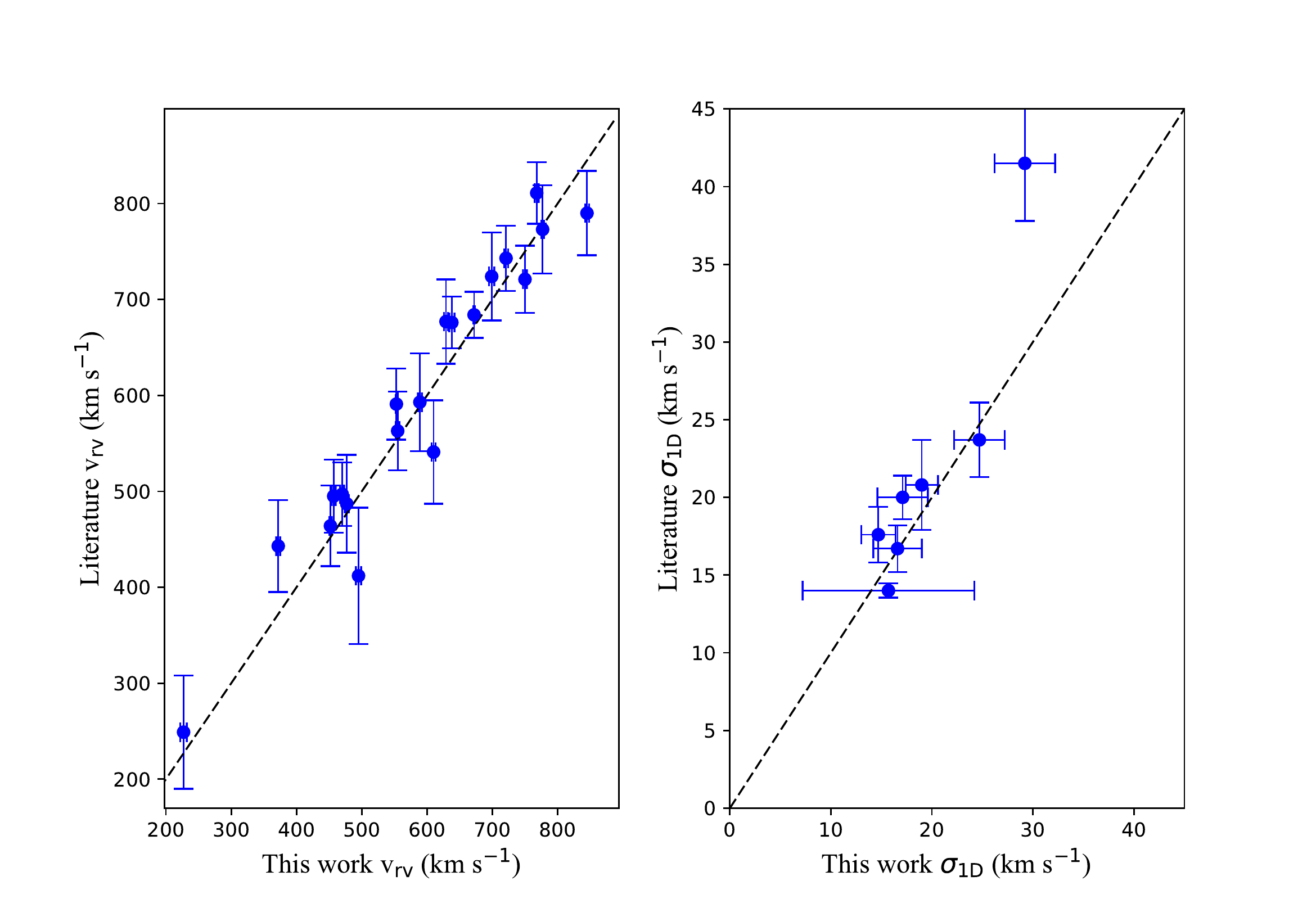}}
      \caption{Left panel: Comparison between radial velocities ($v_{\rm rv}$) measured in this work and in \citet{bea08}. Right panel: Comparison between line-of-sight velocity dispersions inferred in this work and those in the work of \citet{tay10} and \citet{col13}. Dashed lines show the line of equal value. }
         \label{fig:vel}
   \end{figure*}

For completeness, in Figure \ref{fig:vel} we compare our inferred velocities, both radial velocity and line-of-sight velocity dispersion, to values in the literature listed in Table \ref{table:age_fe}. The left panel of Figure \ref{fig:vel} shows excellent agreement between our measured radial velocities with those from the work of \citet{bea08} for all 20 GCs. We note that the errors in the radial velocity measurements are determined from the scatter around the mean value, using the standard deviation of the measurements.\par 
In the right panel of Figure  \ref{fig:vel} we plot our inferred velocity dispersions against those from the work of \citet{tay10} and \citet{col13}. Unfortunately, literature measurements of $\sigma_{\rm 1D}$ are more limited than $v_{\rm rv}$. \citet{tay10}  measured $\sigma_{\rm 1D}$ for seven GCs in our sample. We find good agreement for all but one GC, HGHH-23, where \citet{tay10} finds a significantly higher velocity dispersion. Our value for HGHH-23, however, is in good agreement with that inferred by \citet{rej07}, $\sigma_{\rm 1D}$ = 30.5 $\pm$ 0.2 km s$^{-1}$, well within the errors of our measured velocity dispersion of  $\sigma_{\rm 1D}$ = 29.2 $\pm$ 3.0 km s$^{-1}$. \par
We point out that although our measurements for $\sigma_{\rm 1D}$ agree well with values in the literature, some of our inferred velocity dispersions are relatively high, when compared to the standard $\sigma_{\rm 1D}$ measured for Galactic GCs. From the latest edition of the \citet{har96} catalogue (version 2010), we see that the highest $\sigma_{\rm 1D}$ values measured in Galactic GCs are around $\sim$18 km s$^{-1}$. The velocity dispersion values measured for GCs in NGC 5128 range between 5 $\lesssim \sigma_{\rm 1D} \lesssim$ 30 km s$^{-1}$. That being said, we note that the GC system in NGC 5128 is rich and remarkably larger than that of the Milky Way, and in this study we are probing the brightest GCs in NGC 5128, which could in principle mean that we are looking at the most massive GCs in this early-type galaxy.

\begin{table*}
\caption{Star cluster properties from the Literature.}
\label{table:age_fe}
\centering 
\begin{tabular}{cccccc}
\hline \hline
Cluster & Age$^a$ & [m/H]$^a$  & $v_{\rm rv}$$^b$ & $\sigma_{\rm 1D}$$^c$ & Empirical HB \& AGB$^d$  \\
 & (Gyr) & (dex) & (km s$^{-1}$) & (km s$^{-1}$) & \\
\hline\\
AAT117062 & 10.00 & $-$1.10 &412 $\pm$ 71& - & NGC 0362\\
HGHH-04 & 10.00 & $-$1.63 &724 $\pm$ 46& 14.0 $\pm$ 0.46& NGC 6093 \\
HGHH-06 & 10.00 & $-$1.10 &790 $\pm$ 44& - & NGC 0362\\
HGHH-07 & 11.89 & $-$1.03 &593 $\pm$ 51& - & NGC 0362\\
HGHH-11 & 8.41 & $-$0.43 & 721 $\pm$ 35& 16.7 $\pm$ 1.5& NGC 6838\\
HGHH-17 & 10.00 & $-$0.98 &773 $\pm$ 46& 20.8 $\pm$ 2.9& NGC 0362\\
HGHH-21 & 11.22 & $-$1.10&495 $\pm$ 38& 20.0 $\pm$ 1.4& NGC 0362\\
HGHH-23 & 10.59 & $-$0.48 &684 $\pm$ 24& 41.5 $\pm$ 3.7& NGC 6838\\
HGHH-29 & 11.22 & $-$0.45 &743 $\pm$ 34& 17.6 $\pm$ 1.8& NGC 6838\\
HGHH-34 & 8.41 & $-$0.45&676 $\pm$ 27& - & NGC 6838\\
HGHH-40 & 10.00 & $-$1.25&443 $\pm$ 48& - & NGC 0362\\
HH-080 & 10.00 & $-$1.58&497 $\pm$ 33& - & NGC 6093\\
HH-096 & 10.00 & $-$1.38 & 541 $\pm$ 54& - & NGC 0362\\
HHH86-30 & 5.01 & $-$0.38 &811 $\pm$ 32& - & Palomar 1\\
HHH86-39 & 11.89 & $-$0.48& 464 $\pm$ 42& - & NGC 0362\\
K-029 & 11.89 & $-$0.75& 249 $\pm$ 59& - & NGC 0104\\
K-034 & 11.22 & $-$0.53 &677 $\pm$ 44& - & NGC 0104\\
K-163 & 11.89 & $-$0.98 &487 $\pm$ 51& - & NGC 0362\\
VHH81-03 & 5.31 & $-$0.20& 591 $\pm$ 37& -  & Palomar 1\\
VHH81-05 & 10.00 & $-$1.55  & 563	 $\pm$ 41& - & NGC 6093\\
\hline
\end{tabular}
\begin{minipage}{12cm}~\\
 \textsuperscript{$a$}{Simple stellar population ages and metallicities from \citet{bea08}.}\\
 \textsuperscript{$b$}{Values extracted from the work of \citet{bea08}.}\\
  \textsuperscript{$c$}{Line-of-sight velocity dispersion from \citet{tay10} and \citet{col13}.}\\
  \textsuperscript{$d$}{Galactic GC observations by \citet{sar07} used to represent the HB and AGB stellar stages.}
 \end{minipage}
\end{table*}

\subsection{Element spectral windows}\label{sec:lines}
As detailed in \citet{her17}, one of the main challenges in working with integrated-light observations of star clusters is the degree of blending suffered by the different spectral lines. In \citet{her17} we created optimised spectral windows tailored for individual elements. The windows were carefully selected to mitigate strong blending based on a series of criteria. 
For a more detailed discussion of the optimisation of the spectral windows we refer the reader to \citet{her17} Section 3.4. Briefly, we create two synthetic spectra using the physical parameters of Arcturus as published by \citet{ram11}. One of the spectra is generated excluding all lines for the element in question, whereas the second spectrum includes \textit{only} spectral features for the same element under analysis. Comparing the two spectra we make a pre-selection of spectral lines with a minimal degree of blending based on the following requirements: a) from the normalised spectra the depth of the spectral absorption must have a maximum flux of 0.85, b) neighbouring lines closer than $\pm$0.25 \r{A} are excluded. This pre-selection is done automatically using a Python script. We continue to inspect the recommended element lines and finalise the wavelength windows which include the majority of the clean lines. We note that for Fe and Ti, we select broader windows as most of the wavelength range is populated by spectral lines of these elements. \par

%
\section{Results}\label{results}
\begin{figure}
   \centering
            {\includegraphics[scale=0.53]{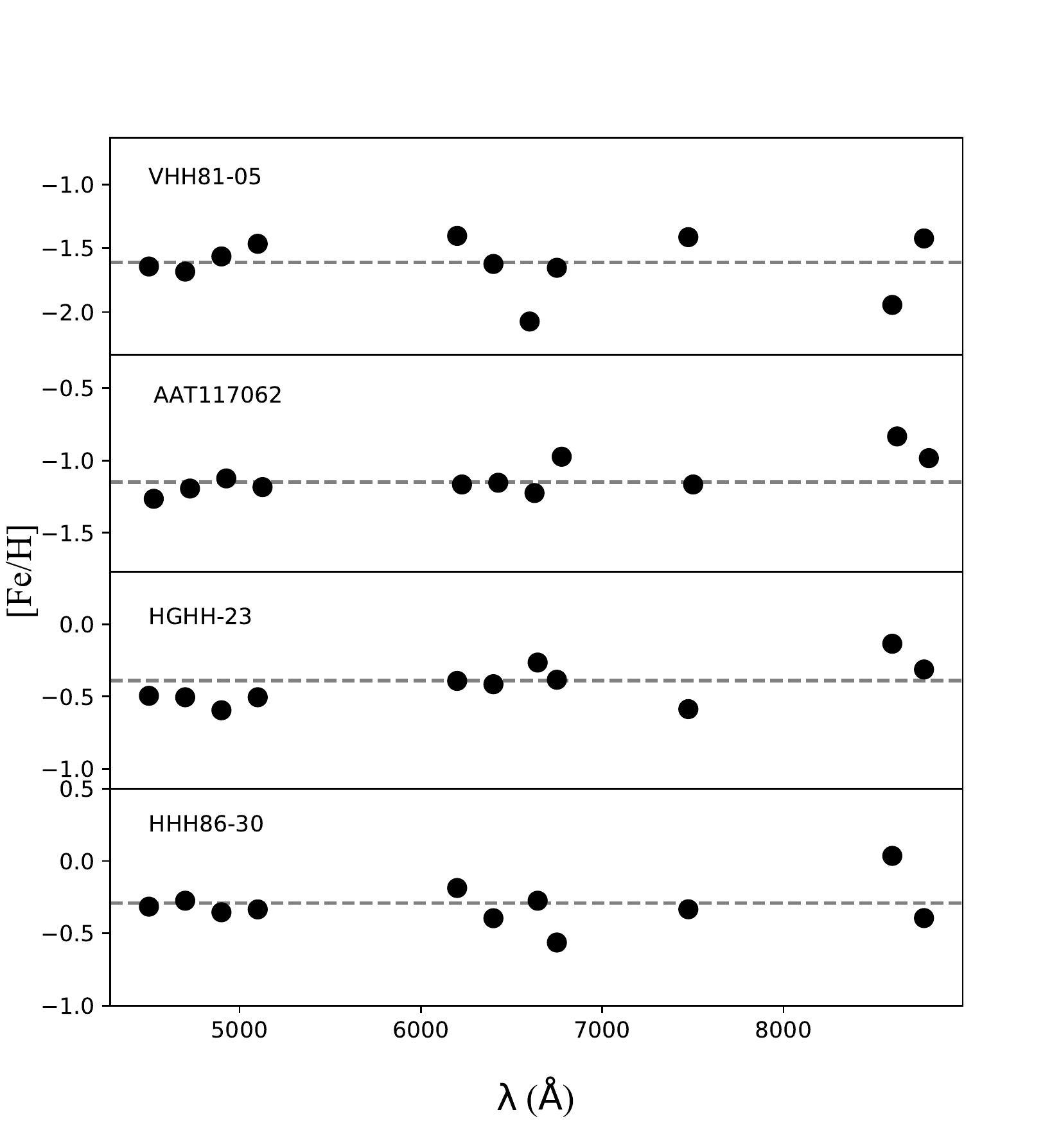}}
      \caption{Fe abundances as a function of wavelength for a selected sample of GCs in NGC 5128. The grey dashed line shows the inferred [Fe/H] for the corresponding clusters. We present the clusters from top to bottom in order of increasing metallicity.}
         \label{fig:fe_trend}
   \end{figure}

After correcting for the radial velocities, and once the smoothing parameter ($\sigma_{\rm sm}$) has been inferred, we fit for the overall metallicity, [Z], keeping $\sigma_{\rm sm}$ fixed. This exercise is performed on each cluster analysing the data in 200 \r{A} intervals. These initial metallicities are only used as a first step, mainly as a scaling factor, before proceeding to measure the individual elements. For wavelength windows of $\geq$100 \r{A}, we match the continua of the model spectra with those of the observations using a cubic spline with three knots. For spectral windows of $<$100 \r{A}, we instead use a first-order polynomial to fit the continuum. We initiate the detailed abundance analysis, first measuring Fe.  We start with those elements with the highest number of lines throughout the spectral coverage. For this element, we again scan the wavelength range analysing 200 \r{A} at a time, covering wavelengths between 4400 \r{A} to 8850 \r{A} excluding the noisy arm edges and telluric contaminated regions. In general we find consistent results throughout the different bins. In Figure \ref{fig:fe_trend} we show the individual Fe abundances for the corresponding wavelength bins for a selected sample. We do not observe any strong trends with wavelength. Additionally, for this same selected sample of GCs, in Figures \ref{fig:mg_app} to \ref{fig:ni_app} we present element abundances as a function of their corresponding wavelength bin. \par

\begin{figure*}
            {\includegraphics[scale=0.60]{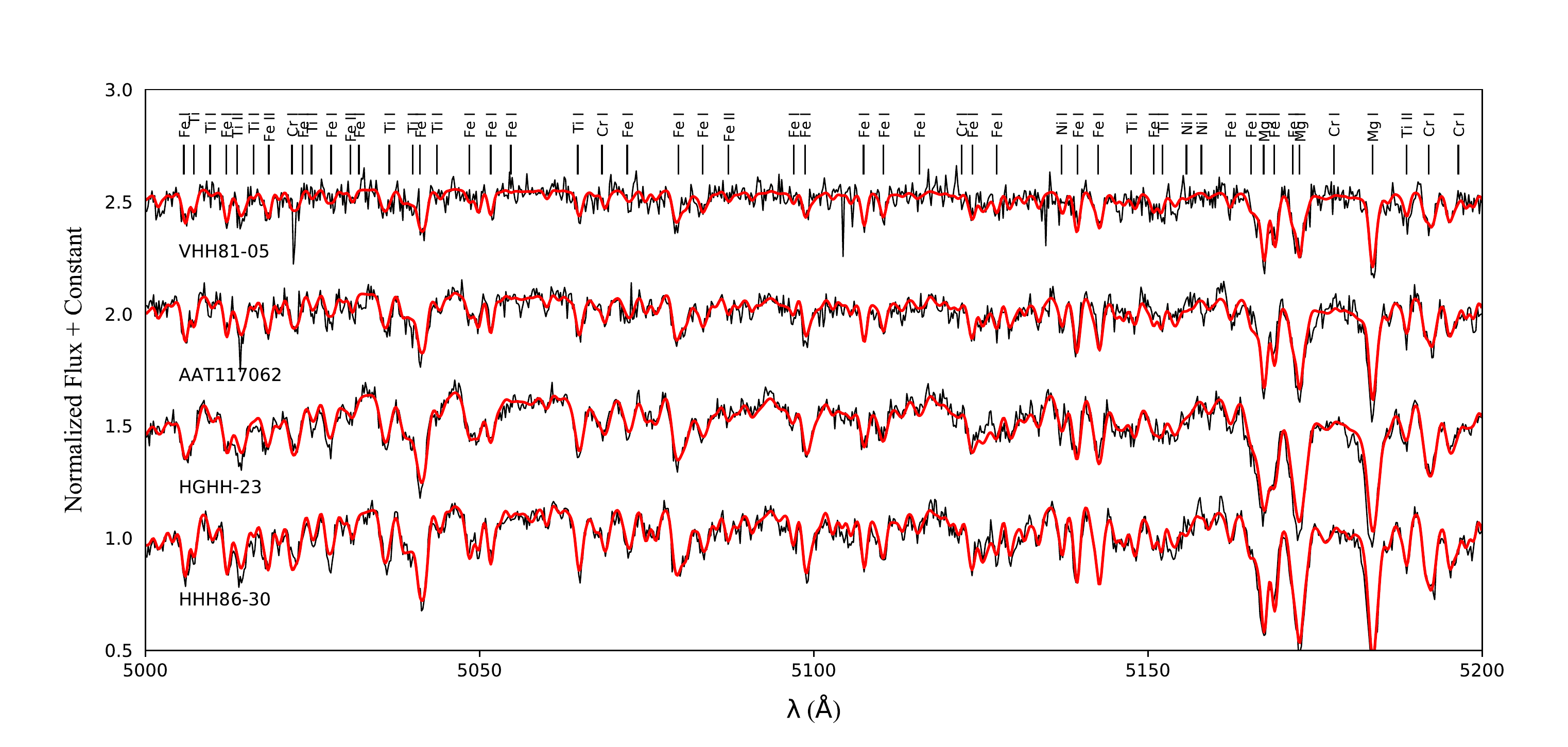}}
      \caption{Normalised synthesis fits for a selected sample of GCs in NGC 5128. In black we show the X-Shooter observations. In red we show the model fit. The individual cluster names are located below the corresponding spectra. We present the clusters from top to bottom in order of increasing metallicity given our results. }
         \label{fig:fits}
   \end{figure*}

We continue down the element list measuring Ti, Ca, Mg, Cr, Mn, Ni, and Na, one element at a time. We point out that the L12 code allows for fittings that include multiple elements simultaneously; however, fitting individual elements appears to be slightly more efficient. For each element we use the tailored wavelength windows optimised following the criteria presented in Section \ref{sec:lines}. As we proceed with the analysis, we keep the measured abundances fixed to their best fitting value, and continue with the rest of the elements. In Tables \ref{tab:1} to \ref{tab:20} we show the elements, wavelength bins, best-fitting abundances, and their respective 1-$\sigma$ uncertainties from the $\chi^{2}$ fit. Additionally, in Figure \ref{fig:fits} we show normalised synthesis fits for a selected star cluster sample covering wavelengths 5000-5200 \r{A}. \par
In Tables \ref{table:derived_metal} and \ref{table:derived_prop} we present our final abundances, along with their corresponding errors, $\sigma_{X}$. The values in these tables are estimates of the weighted average abundances. When calculating these values we assume weights defined as $w_{i}= 1/\sigma_{i}^{2}$, where $\sigma_{i}$ represents the 1-$\sigma$ errors listed in Tables \ref{tab:1} to \ref{tab:20}. Given that the scatter in the individual measurements is larger than the errors based on the $\chi^{2}$ fits, we believe that the standard deviation, $\sigma_{\rm STD}$, appears to be more representative of the actual uncertainties. Therefore, we turn the $\sigma_{\rm STD}$ into formal errors of the mean abundances following:

\begin{equation}\label{eq:errors}
\sigma_{X} = \frac{\sigma_{\rm STD}}{\sqrt{N - 1}}
\end{equation}
In equation \ref{eq:errors}, $\sigma_{\rm STD}$ and $N$ represent the standard deviation and number of bins for the elements in question, respectively.

 \subsection{Sensitivity to input isochrones}\label{sen_input}
The integrated-light analysis done here relies on theoretical models. We make use of $\alpha$-enhanced isochrones by the Dartmouth group, as we are mainly studying relatively older populations. However, to test the uncertainties involved in a particular choice of theoretical isochrones, we repeat the full abundance analysis using instead solar-scaled PARSEC isochrones \citep{bre12}. An additional difference between the $\alpha$-enhanced Dartmouth isochrones and the solar-scaled PARSEC models, other than the chemical composition, is the stellar evolutionary phase coverage. The PARSEC models include the HB and AGB phases and do not require empirical data for the post-RGB stellar phases.\par
\begin{figure}
            {\includegraphics[scale=0.55]{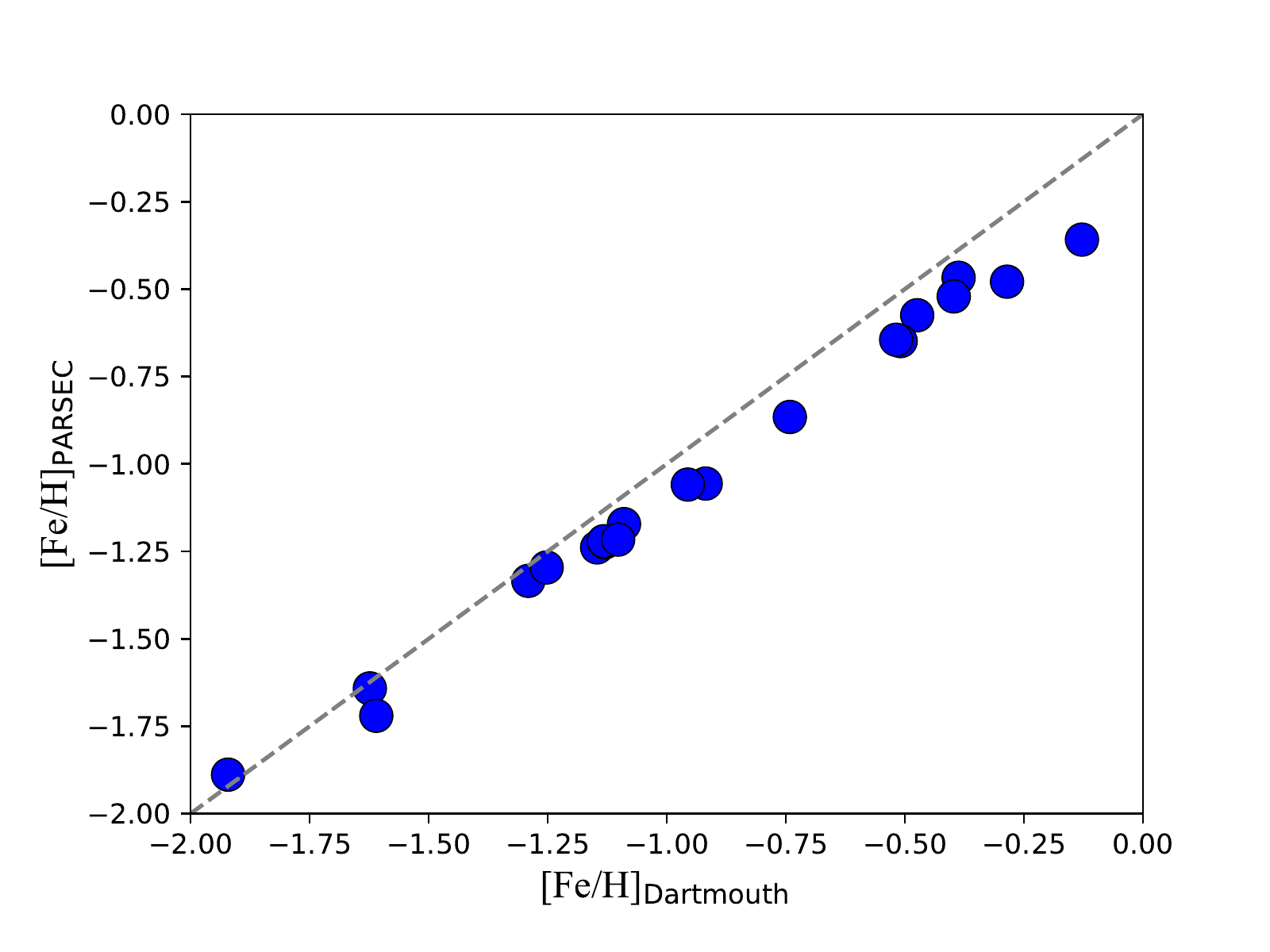}}
      \caption{Comparison between the metallicities inferred using $\alpha$-enhanced isochrone (Dartmouth), and solar-scaled isochrones (PARSEC). The grey dashed line shows equal values.  }
         \label{fig:fe_mod}
   \end{figure}

 
In Figure \ref{fig:fe_mod} we compare [Fe/H] as measured using $\alpha$-enhanced isochrones (Dartmouth) and solar-scaled isochrones (PARSEC). We find that the average difference is $\lesssim$0.1 dex. The highest difference is observed in the most metal-rich GC, VHH81-03, with a difference in the estimated [Fe/H] of $\sim$0.2 dex. From Figure \ref{fig:fe_mod}  we can see that at low metallicities ([Fe/H] < $-$1.0), the inferred [Fe/H] abundances agree with each other. \citet{lar17} compared metallicities obtained from integrated-light observations for 7 Galactic GCs using Dartmouth isochrones and MIST \citep[MESA Isochrones and Stellar Tracks, ][]{cho16, dot16} models. Their sample spanned metallicities from [Fe/H] $\sim-$2.4 to [Fe/H] $\sim-$0.5 dex. Similar to our test results, \citet{lar17} found comparable metallicities and abundances for the most metal-poor clusters, [Fe/H] $<-$2.0 using both models, Dartmouth and MIST. The metal-rich regime appeared to have larger model dependencies than what was observed for the metal-poor end. Similar results were observed by \citet{col16} when comparing their Galactic GC [Fe/H] derived from integrated-light observations and those from \citet[][2010 edition]{har91}. They find accurate values to within $\sim$0.1 dex at low metallicities, [Fe/H]$\lesssim-$0.3 dex, and observe higher offsets for the metal-rich clusters. This might be caused by the similarities between different theoretical isochrones, especially in what appears to be the better understood metal-poor regime. In general, it seems that Fe abundances from integrated-light analysis are more robust at lower metallicities, than at higher values ([Fe/H]$> -$1.0 dex). \par

\begin{figure}
            {\includegraphics[scale=0.45]{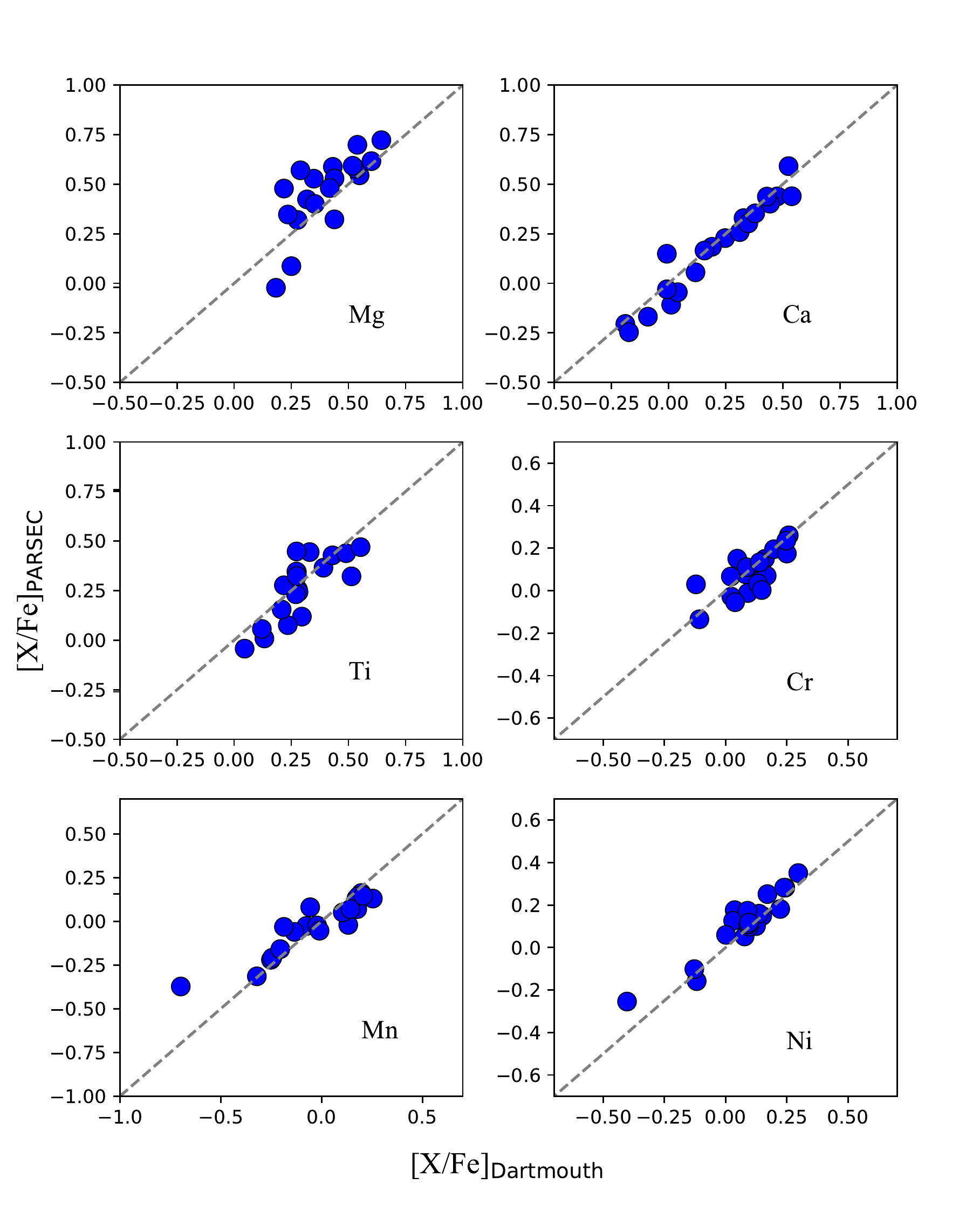}}
      \caption{Comparison between the detailed abundances measured using $\alpha$-enhanced isochrone (Dartmouth), and solar-scaled isochrones (PARSEC). The element in question is indicated in the corresponding panel. The grey dashed line shows equal values. }
         \label{fig:abun_mod}
   \end{figure}

In the different panels of Figure \ref{fig:abun_mod} we compare the abundances obtained for the rest of the elements. Replacing the  $\alpha$-enhanced isochones with solar-scaled models increases the [Mg/Fe] abundances on average by $\sim$0.12 dex. For [Ca/Fe] ratios the differences between the two runs are comparatively smaller, with average differences of the order of 0.05 dex (top right panel in Figure \ref{fig:abun_mod}). Changing the isochrone models from $\alpha$-enhanced to solar-scaled decreases the [Ti/Fe] ratios only slightly, with average differences of $\sim$0.08 dex. Overall, we find no correlation between the differences in the abundance ratios of [(Ca, Ti)/Fe] and the [Fe/H] abundances. For Mg, however, we see a slight correlation where the largest differences are seen in the two clusters with higher metallicities, VHH81-03 and HHH86-30.\par
In addition to the $\alpha$-elements, we also estimate abundances for Fe-peak elements changing the input models. The average differences for [Cr/Fe], [Mn/Fe], and [Ni/Fe] ratios is $\sim$0.05, 0.08, and 0.05, respectively. Comparison abundances for Fe-peak elements are also shown in Figure \ref{fig:abun_mod}. No general correlations between the ratio differences and metallicity are observed for any of the Fe-peak elements. \par

 \subsection{Fe and Ca abundances: comparison with literature}
\begin{figure}
            {\includegraphics[scale=0.42]{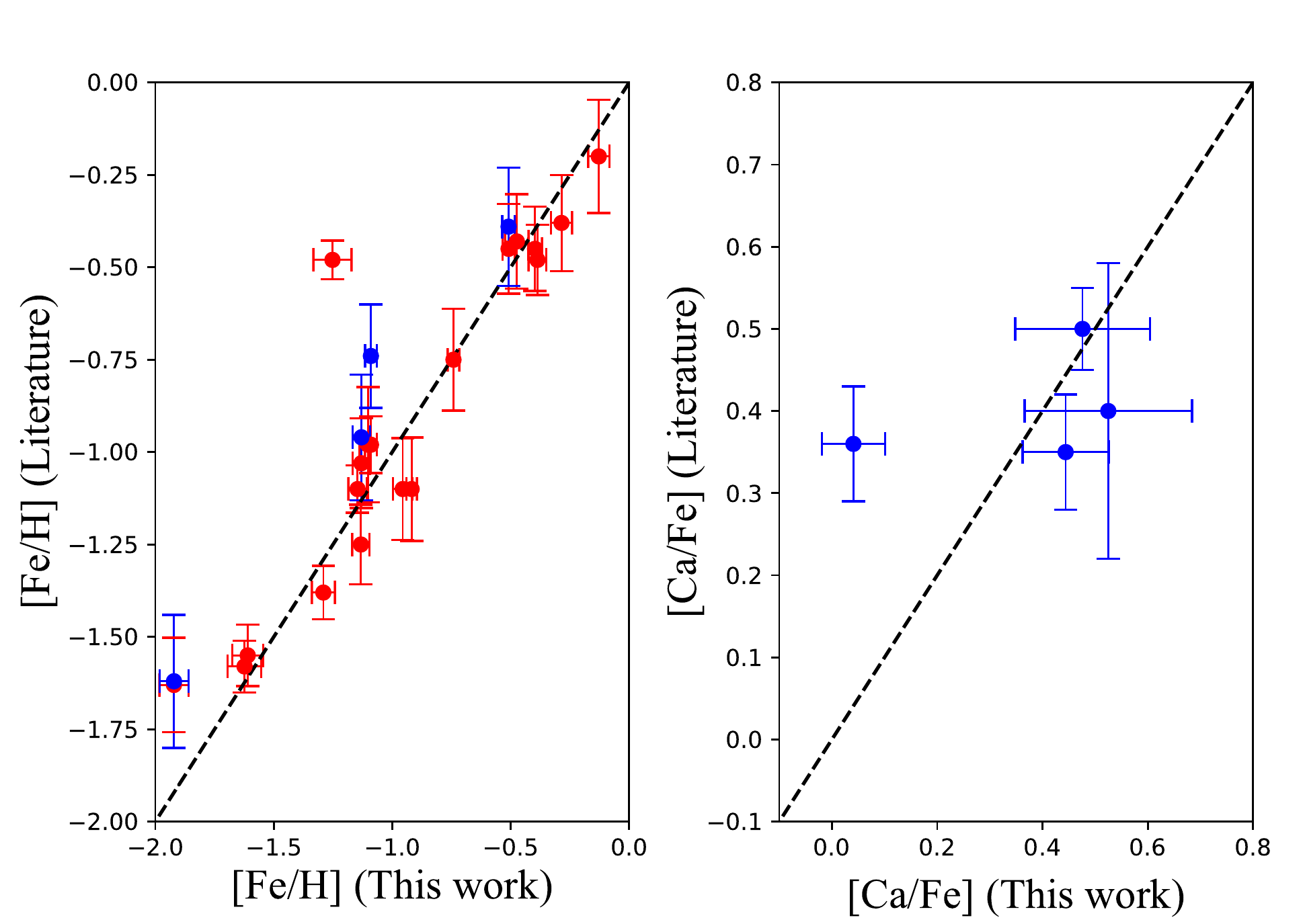}}
      \caption{[Fe/H] and [Ca/Fe] abundance comparison between this work and that from \citealt{col13} (in blue circles), and \citealt{bea08} (in red circles). The dash line in both panels represents a 1:1 relation. }
         \label{fig:col}
   \end{figure}

\citet{bea08} derived empirical metallicities, [m/H], for all 20 GCs in our sample. In the left panel of Figure \ref{fig:col} (red circles) we compare our metallicities to those measured by \citet{bea08}. In general, we find excellent agreement between their work, and our inferred [Fe/H] abundances, except for GC HHH86-39. For this cluster \citeauthor{bea08} estimate a metallicity of [m/H] = $-$0.48, whereas we find an [Fe/H] abundance of $-$1.29 $\pm$ 0.05 dex. As described in Section \ref{theo} for the initial selection of isochrones and empirical CMDs we adopt the ages and metallicities by  \citet{bea08}. For HHH86-39 we initially select an isochrone with metallicity $-$0.48 dex, and see that the inferred metallicities converge onto lower values. We continue re-estimating the metallicity for HHH86-39 using instead an isochrone with [Fe/H] $=-$1.28 dex and estimate a consistent [Fe/H] abundance of $-$1.29 $\pm$ 0.05. Given the self-consistency between the input isochrone and the measured [Fe/H] abundance, we continue our analysis adopting the inferred [Fe/H] $=-$1.29 $\pm$ 0.05 dex. As we noted earlier, the line index techniques used by \citet{bea08} do not measure [Fe/H] necessarily. However, an overall metallicity comparison is still a useful test. \par
Using high-resolution and high signal-to-noise integrated-light observations \citet{col13} measured Fe and Ca abundances for a sample of 10 GCs in NGC 5128. Of the 10 clusters in \citet{col13} we have 4 clusters in common between their study and our work. From a direct comparison to the metallicities obtained by \citet{col13} we see an average offset of $\sim$$+$0.2 dex between their metallicities and those estimated here; however, they all agree within 2$\sigma$ (See Figure \ref{fig:col} left panel). \par
 Similarly, in the right panel of Figure \ref{fig:col} we compare our inferred [Ca/Fe] ratios to those obtained by \citet{col13}. The [Ca/Fe] ratios of three out of four clusters are in excellent agreement, within 1$\sigma$, to those measured in the high-resolution observations.  The Ca abundance for the fourth cluster, HGHH-29, appears to be a 3-$\sigma$ outlier compared to our [Ca/Fe] measurements with an offset of the order of $\sim$0.3 dex. We visually inspect the Ca model fits and find no anomalies, as well as  consistent [Ca/Fe] abundances between the different bins. \par 

\begin{table}
\caption{Derived velocities. }
\label{table:derived_vel}
\centering 
\begin{tabular}{ccc}
\hline \hline
Cluster & $v_{\rm rv}$  & $\sigma_{\rm 1D}$ \\
 & (km s$^{-1}$) & (km s$^{-1}$)\\
\hline\\
AAT117062 & 495 $\pm$ 4& 20.2 $\pm$ 4.4 \\
HGHH-04 & 699 $\pm$ 4& 15.7 $\pm$ 8.5\\
HGHH-06 & 845 $\pm$ 3& 28.5 $\pm$ 3.7\\
HGHH-07 & 589 $\pm$ 3& 24.7 $\pm$ 2.5\\
HGHH-11 & 750 $\pm$ 3& 16.6 $\pm$ 2.4\\
HGHH-17 & 777 $\pm$ 2 & 19.0 $\pm$ 1.6\\
HGHH-21 & 457 $\pm$ 3& 17.1 $\pm$ 2.5\\
HGHH-23 & 672 $\pm$ 1& 29.2 $\pm$ 3.0\\
HGHH-29 & 721 $\pm$ 3& 14.7 $\pm$ 1.7\\
HGHH-34 & 638 $\pm$ 4& 10.8 $\pm$ 2.8\\
HGHH-40 & 372 $\pm$ 3& 4.7 $\pm$ 5.1\\
HH-080 & 470 $\pm$ 3& 12.4 $\pm$ 4.0\\
HH-096 & 610 $\pm$ 3& 13.6 $\pm$ 5.0\\
HHH86-30 & 768 $\pm$ 3& 22.3 $\pm$ 1.7\\
HHH86-39 & 227 $\pm$ 5& 12.5 $\pm$ 2.2\\
K-029 & 629 $\pm$ 3& 25.3 $\pm$ 1.1\\
K-034 & 452 $\pm$ 2& 14.4 $\pm$ 2.1\\
K-163 & 477 $\pm$ 3& 22.2 $\pm$ 1.3\\
VHH81-03 & 553 $\pm$ 1& 17.5 $\pm$ 2.4\\
VHH81-05 & 555 $\pm$ 3& 10.2 $\pm$ 5.3\\
\hline
\end{tabular}
\end{table}

\begin{table}
\caption{Derived metallicities. }
\label{table:derived_metal}
\centering 
\begin{tabular}{cc}
\hline \hline
Cluster & [Fe/H] $\pm\;\sigma_{\rm Fe}$\\
 & (dex) \\
\hline\\
AAT117062  &$-$1.15 $\pm$ 0.04\\
HGHH-04 & $-$1.92 $\pm$ 0.06\\
HGHH-06 &$-$0.92 $\pm$ 0.02\\
HGHH-07&$-$1.13 $\pm$ 0.03\\
HGHH-11 &$-$0.47 $\pm$ 0.05\\
HGHH-17 &$-$1.09 $\pm$ 0.03\\
HGHH-21 &$-$0.96 $\pm$ 0.04\\
HGHH-23 &$-$0.39 $\pm$ 0.04\\
HGHH-29 &$-$0.51 $\pm$ 0.03\\
HGHH-34 &$-$0.40 $\pm$ 0.03\\
HGHH-40 &$-$1.13 $\pm$ 0.04\\
HH-080 &$-$1.62 $\pm$ 0.07\\
HH-096 &$-$1.29 $\pm$ 0.05\\
HHH86-30&$-$0.29 $\pm$ 0.04\\
HHH86-39 &$-$1.25 $\pm$ 0.08\\
K-029 &$-$0.74 $\pm$ 0.02\\
K-034 &$-$0.52 $\pm$ 0.03\\
K-163 &$-$1.10 $\pm$ 0.04\\
VHH81-03 &$-$0.13 $\pm$ 0.05\\
VHH81-05 &$-$1.61 $\pm$ 0.07\\
\hline
\end{tabular}
\end{table}

\begin{table*}
\caption{Derived abundances. }
\label{table:derived_prop}
\centering 
\begin{tabular}{cccccccc}
\hline \hline
Cluster &  [Mg/Fe] $\pm\;\sigma_{\rm Mg}$ & [Ca/Fe] $\pm\;\sigma_{\rm Ca}$& [Ti/Fe] $\pm\;\sigma_{\rm Ti}$& [Na/Fe] $\pm\;\sigma_{\rm Na}$&[Cr/Fe] $\pm\;\sigma_{\rm Cr}$& [Mn/Fe] $\pm\;\sigma_{\rm Mn}$& [Ni/Fe] $\pm\;\sigma_{\rm Ni}$\\
 & (dex) & (dex) & (dex) & (dex) & (dex) & (dex) & (dex) \\
\hline\\
AAT117062  &$+$0.55 $\pm$ 0.13&$+$0.25 $\pm$ 0.10&$+$0.39 $\pm$ 0.14& - & $+$0.11 $\pm$ 0.20&$-$0.08 $\pm$ 0.24&$+$0.17 $\pm$ 0.07\\
HGHH-04 &$+$0.44 $\pm$ 0.17&$+$0.53 $\pm$ 0.16&$+$0.51 $\pm$ 0.10& - &$+$0.12 $\pm$ 0.20&$-$0.70 $\pm$ 0.29&$+$0.04 $\pm$ 0.11\\
HGHH-06 &$+$0.64 $\pm$ 0.06&$+$0.33 $\pm$ 0.16&$+$0.33 $\pm$ 0.10& - &$+$0.16 $\pm$ 0.13&$-$0.25 $\pm$ 0.29&$+$0.24 $\pm$ 0.15\\
HGHH-07&$+$0.53 $\pm$ 0.13&$+$0.48 $\pm$ 0.13&$+$0.28 $\pm$ 0.10&$+$0.26 $\pm$ 0.20&$+$0.08 $\pm$ 0.19&$-$0.05 $\pm$ 0.28&$+$0.13 $\pm$ 0.08\\
HGHH-11 &$+$0.32 $\pm$ 0.11&$-$0.09 $\pm$ 0.12&$+$0.13 $\pm$ 0.15&$+$0.04 $\pm$ 0.08&$-$0.12 $\pm$ 0.08&$+$0.25 $\pm$ 0.29&$+$0.08 $\pm$ 0.11\\
HGHH-17 &$+$0.52 $\pm$ 0.10&$+$0.44 $\pm$ 0.08&$+$0.27 $\pm$ 0.05&$+$0.04 $\pm$ 0.18&$+$0.17 $\pm$ 0.30&$-$0.02 $\pm$ 0.17&$+$0.15 $\pm$ 0.07\\
HGHH-21 &$+$0.54 $\pm$ 0.09&$+$0.31 $\pm$ 0.09&$+$0.43 $\pm$ 0.11& - &$+$0.20 $\pm$ 0.24&$+$0.17 $\pm$ 0.07&$+$0.24 $\pm$ 0.19\\
HGHH-23 &$+$0.28 $\pm$ 0.14&$+$0.01 $\pm$ 0.11&$+$0.05 $\pm$ 0.23&$+$0.38 $\pm$ 0.08&$+$0.09 $\pm$ 0.19&$+$0.13 $\pm$ 0.29&$-$0.12 $\pm$ 0.18\\
HGHH-29 &$+$0.43 $\pm$ 0.06&$+$0.04 $\pm$ 0.06&$+$0.28 $\pm$ 0.07& - &$+$0.26 $\pm$ 0.29&$+$0.20 $\pm$ 0.23&$+$0.12 $\pm$ 0.14\\
HGHH-34 &$+$0.35 $\pm$ 0.09&$+$0.00 $\pm$ 0.06&$+$0.49 $\pm$ 0.18& - &$+$0.02 $\pm$ 0.18&$+$0.18 $\pm$ 0.06&$+$0.23 $\pm$ 0.10\\
HGHH-40 &$+$0.24 $\pm$ 0.09&$+$0.19 $\pm$ 0.07&$+$0.27 $\pm$ 0.25& - &$+$0.05 $\pm$ 0.44&$-$0.13 $\pm$ 0.06&$+$0.30 $\pm$ 0.32\\
HH-080 &$+$0.18 $\pm$ 0.14&$+$0.35 $\pm$ 0.09&$+$0.30 $\pm$ 0.14& - &$+$0.04 $\pm$ 0.19&$-$0.01 $\pm$ 0.18&$+$0.03 $\pm$ 0.14\\
HH-096 &$+$0.35 $\pm$ 0.22&$+$0.38 $\pm$ 0.05&$+$0.27 $\pm$ 0.07& - &$+$0.09 $\pm$ 0.12&$-$0.24 $\pm$ 0.22&$+$0.14 $\pm$ 0.10\\
HHH86-30&$+$0.29 $\pm$ 0.06&$-$0.19 $\pm$ 0.13&$+$0.27 $\pm$ 0.12&$+$0.26 $\pm$ 0.10&$+$0.25 $\pm$ 0.17&$+$0.18 $\pm$ 0.08&$+$0.10 $\pm$ 0.07\\
HHH86-39 &-&$+$0.12 $\pm$ 0.20&$+$0.24 $\pm$ 0.25& $-$0.05 $\pm$ 0.11 &$+$0.13 $\pm$ 0.63&$-$0.32 $\pm$ 0.26&$-$0.13 $\pm$ 0.15\\
K-029 &$+$0.44 $\pm$ 0.11&$+$0.16 $\pm$ 0.14&$+$0.22 $\pm$ 0.12&$+$0.70 $\pm$ 0.08&$+$0.15 $\pm$ 0.25&$+$0.11 $\pm$ 0.50&$+$0.00 $\pm$ 0.11\\
K-034 &$+$0.42 $\pm$ 0.12&$-$0.01 $\pm$ 0.08&$+$0.21 $\pm$ 0.06&$+$0.34 $\pm$ 0.08&$+$0.25 $\pm$ 0.19&$+$0.14 $\pm$ 0.03&$+$0.09 $\pm$ 0.11\\
K-163 &$+$0.60 $\pm$ 0.13&$+$0.43 $\pm$ 0.09&$+$0.28 $\pm$ 0.06&$+$0.65 $\pm$ 0.22&$+$0.02 $\pm$ 0.19&$-$0.21 $\pm$ 0.17&$+$0.09 $\pm$ 0.13\\
VHH81-03 &$+$0.22 $\pm$ 0.08&$-$0.17 $\pm$ 0.16&$+$0.12 $\pm$ 0.10& - &$+$0.14 $\pm$ 0.29&$+$0.21 $\pm$ 0.29&$+$0.10 $\pm$ 0.10\\
VHH81-05 &$+$0.25 $\pm$ 0.12&$+$0.54 $\pm$ 0.09&$+$0.55 $\pm$ 0.17& - &$-$0.11 $\pm$ 0.61&$-$0.19 $\pm$ 0.06&$-$0.40 $\pm$ 0.11\\
\hline
\end{tabular}
\end{table*}

\section{Discussion}\label{discuss}
In this section we discuss our abundance measurements, along with a comparison to other NGC 5128 abundance studies in the literature. We also compare the abundance patterns observed in NGC 5128 to those studied in different environments, such as the MW and M 31.

 \subsection{$\alpha$-elements}
 As mentioned before, it is believed that $\alpha$-elements (O, Mg, Si, Ca, and Ti) are primarily produced in high-mass stars and ejected through core-collapse supernovae \citep{woo95}. Given that these elements are believed to be part of a homogenous group, it is common for abundance studies to average their abundances to obtain a single [$\alpha$/Fe] ratio. In general, the abundances of these different elements appear to correlate tightly with each other in Galactic stars. However, [Mg/Fe] has been observed to be slightly more enhanced than the rest of the $\alpha$-elements in dSph galaxies \citep{she04,ven04}. This difference between the Mg ratios and those of Ca and Ti could have nucleosynthetic origins. Although all three elements, Mg, Ca, and Ti, are produced inside high-mass stars, Mg is created through hydrostatic C- and O-burning, in contrast to Ca and Ti isotopes which are formed in the $\alpha$-process \citep{woo95, nak01}. If these two processes do not occur together, then one would expect to see differences in the [Mg/Fe] ratios when compared to those of [Ca/Fe] and [Ti/Fe]. This being said, the differences in the [Mg/Fe] and [(Ca,Ti)/Fe] ratios might also be explained by different star formation histories, mixing timescales or IMFs.  \par
 
\begin{figure*}
            {\includegraphics[scale=0.7]{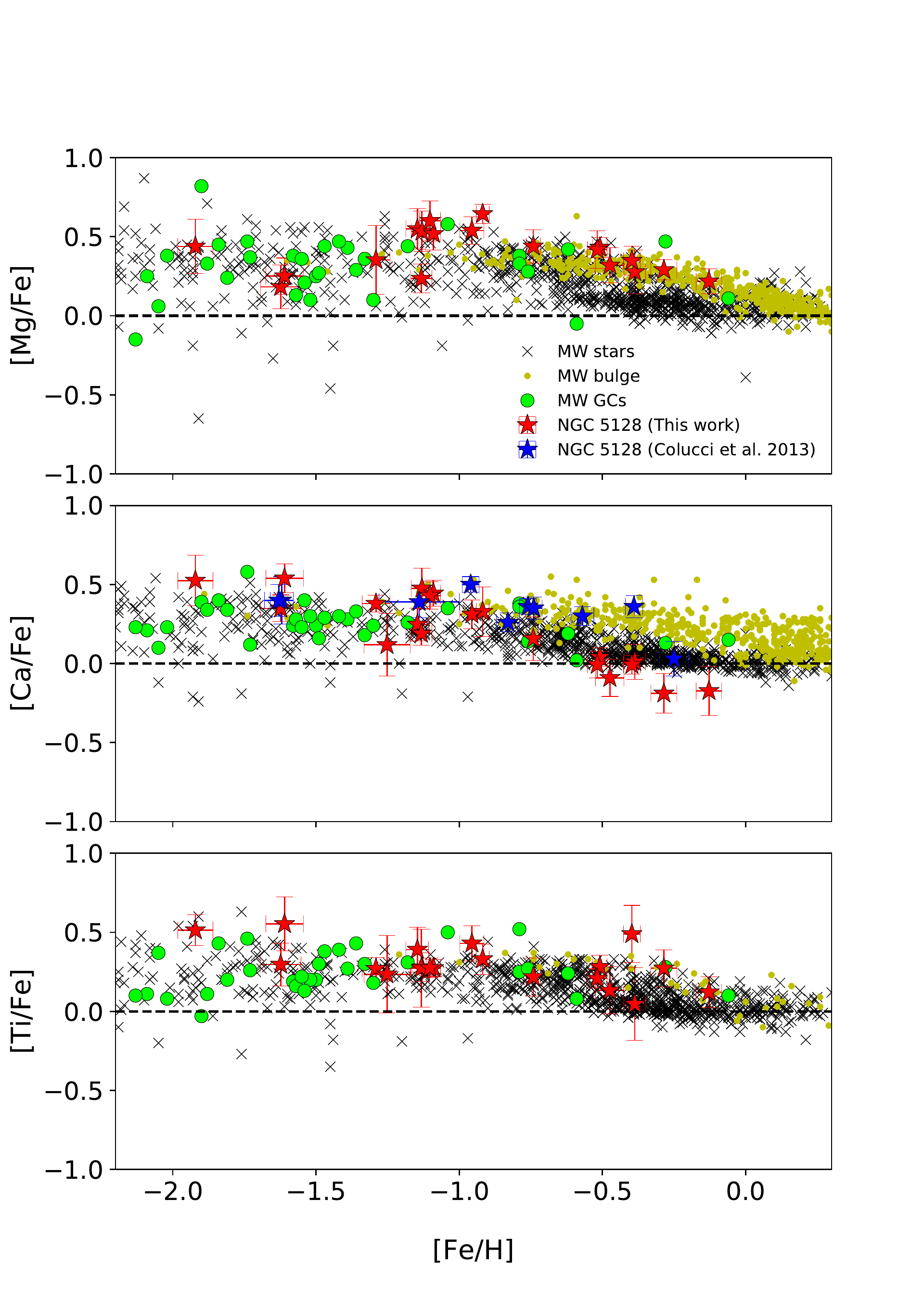}}
      \caption{Top: [Mg/Fe] vs. [Fe/H]. Middle: [Ca/Fe] vs. [Fe/H]. Bottom:  [Ti/Fe] vs. [Fe/H]. In red stars we show the $\alpha$ abundances for GCs in NGC 5128 (this work). Blue stars show the abundances for GCs in NGC 5128 by \citet{col13}. Green circles show abundance measurements of GCs in the MW by \citet{pri05}. Black crosses show the ratio abundance of MW stars found in the literature \citep{red03, red06, ven04,ish13}. Abundance measurements for bulge stars in the MW are shown in yellow points \citep{ben13, jon14, gon15}.}
         \label{fig:alpha}
   \end{figure*}

In Figure \ref{fig:alpha} we display the [Mg/Fe], [Ca/Fe], and [Ti/Fe] ratios as a function of metallicity for all 20 GCs in NGC 5128 (red stars). We compare our NGC 5128 $\alpha$-element abundances to those observed in Galactic stars for both halo and disk stars (black crosses), MW bulge stars (yellow points), and GCs in the MW (green circles). We also include the [Ca/Fe] abundances for the GC sample in NGC 5128 studied by \citet{col13} as blue stars. It is important to note that the MW stellar abundances shown as black crosses in Figure \ref{fig:alpha} cover the different Galactic components, halo, thin and thick disk. An analysis focused on the kinematical information available for different MW stars allowed \citet{ven04} to assign stars to different Galactic components finding that halo stars reach metallicities as high as [Fe/H]$\sim-$1 dex. At higher metallicities than this MW disk stars seem to predominate. As seen from Figure \ref{fig:alpha}, all of the star clusters in our sample of NGC 5128 with [Fe/H] < $-$0.5 dex are enhanced in [$\alpha$/Fe] relative to solar abundance.  
 
It is clear from the top panel of Figure \ref{fig:alpha} that we measure relatively high enhancement in [Mg/Fe] ratios, when compared to what is seen in the MW (GCs and field stars). For lower metallicities, [Fe/H] < $-$0.75 dex, we measure an average [Mg/Fe] = $+$0.49 $\pm$ 0.05 for the GCs in NGC 5128, compared to [Mg/Fe] = $+$0.33 $\pm$ 0.01 for Galactic stars (mainly halo stars), and [Mg/Fe] = $+$0.33 $\pm$ 0.03 for GCs in the MW. We note that the errors are estimated by turning the standard deviation estimates into errors on the mean abundances. The highest enhancement in the [Mg/Fe] ratios is observed in 6 GCs with metallicities in the range of $-$1.15 $<$ [Fe/H] $<$ $-$0.92 dex, with average [Mg/Fe] $\sim$ 0.56 dex. If we instead compare the mean [Mg/Fe] ratios for GCs with metallicities [Fe/H] $<-$1.25 dex, we find [Mg/Fe] = $+$0.31 $\pm$ 0.06 and $+$0.32 $\pm$ 0.04 for GCs in NGC 5128 and the MW, respectively. To confirm if the enhancement at metallicities [Fe/H] <$-$0.75 dex is genuine we recommend extending the GCs sample size in NGC 5128 to cover a larger range in metallicities, similar to the GC sample in the MW (with metallicities at the lower end extending to [Fe/H] $\sim-$2.4 dex). For metallicities higher than [Fe/H] > $-$0.75, the [Mg/Fe] ratios are close to the upper envelope of measurements of field stars in the MW (black crosses) and rather comparable to average [Mg/Fe] values found in Galactic bulge stars (yellow points). 
For comparison to the GCs in the MW, we take the weighted averages and find mean [Mg/Fe] ratios of $+0.34\pm0.03$ and $+0.24\pm0.13$ for the GCs in NGC 5128 and Galactic GCs, respectively, where we might be detecting a slightly higher enhancement in the NGC 5128 GCs compared to the few GCs in the MW at these metallicities. \par

For the Ca abundances in GCs in NGC 5128, \citet{col13} reported a relatively high enhancement of [Ca/Fe] = $+$0.37 $\pm$ 0.07 for clusters with metallicities of [Fe/H] < $-$0.4. For consistency we take the [Ca/Fe] ratios of \citet{col13} and calculate a simple mean for clusters with  [Fe/H] < $-$0.75, which gives [Ca/Fe] = $+$0.39 $\pm$ 0.03. Taking the mean of our inferred [Ca/Fe] for GCs with [Fe/H] < $-$0.75, we obtain [Ca/Fe] = $+$0.36 $\pm$ 0.04, certainly within the errors of the mean abundance ratio of \citet{col13}. Again, these values are higher than what is observed in the MW stars at the same metallicities, [Ca/Fe] $+$0.27 $\pm$ 0.01. The agreement between the two independent studies of NGC 5128 confirm the Ca abundance enhancement in the metal-poor GCs. At metallicities [Fe/H]>$-$0.75, the [Ca/Fe] ratios decrease to solar/sub-solar abundances, which is rarely observed in any of the MW components. The two most metal-rich GCs in our sample, HHH86-30 and VHH81-03, appear to be those with the lowest [Ca/Fe] abundance ratios. We note that abundances in the GC sample by \citet{col13} show instead a less steep decrease in [Ca/Fe] towards higher metallicities.  \par

We also see enhanced Ti abundances with a mean value [Ti/Fe] = $+$0.32  $\pm$ 0.03, for metallicities [Fe/H] < $-$0.75 (see bottom panel of Figure \ref{fig:alpha}). This in contrast to the mean [Ti/Fe] ratio seen in Galactic stars, [Ti/Fe] = $+$0.25 $\pm$ 0.01, and a similar mean average for MW GCs, [Ti/Fe] = $+$0.25 $\pm$ 0.02. Resembling what we observe in the [Mg/Fe] ratios at higher metallicities, [Fe/H]$>-$0.75, we find that the [Ti/Fe] abundances are more comparable to the few [Ti/Fe] ratios measured in the the MW bulge, which tend to be on the upper envelope of the stellar abundances in the MW (black crosses in Figure \ref{fig:alpha}). \par

In general, the [$\alpha$/Fe] abundances for GCs with [Fe/H] < $-$0.75 in NGC 5128 appear to be higher than those observed in the MW by $\sim$0.10-0.15 dex. We note that systematic differences between the measurement techniques can not be ruled out as a possible cause for the observed enhancement of the  [$\alpha$/Fe] ratios. A broad range of studies infer different abundances provided the different methods, models and parameters used. As discussed in Section \ref{sen_input}, similar offsets are seen when changing the input isochrones from $\alpha$-enhanced to solar-scaled for Mg abundances. However, the agreement between the Ca enhancement in the metal-poor GCs in the independent study by \citet{col13} and this work adds confidence to the validity of the measurements. \par

If the overall enhancement in the $\alpha$-elements measured in this work is genuine, this could hint at an IMF skewed to high-mass stars. We also note that in addition to relatively high $\alpha$ abundances, through an empirical spectroscopic study of the metallicity distribution function  of the GC system in NGC 5128 \citet{bea08} find that the mean metallicity of this galaxy is $\sim$0.5 dex more metal-rich than that of the MW GC system \citep{sch05, puz02}. The combination of these two effects, high average metallicity and enhanced [$\alpha$/Fe] ratios, might indicate that high-mass SNe II have been the main drivers in chemically diversifying the galaxy environment \citep{mc97}. In a large study of metallicity and abundance ratios of early-type galaxies such as NGC 5128, using galaxy chemical evolution semi-analytical models (SAMs) with feedback from active galactic nuclei (AGN) \citet{arr10} find that in order to reproduce the galaxy observations, more specifically the observed positive slope in the Mass-[$\alpha$/Fe] relation, they have to apply a mildly top-heavy IMF along with a lower fraction of binaries that explodes as SNe Ia. The original expectation of this work was that the inclusion of the AGN feedback in the simulations would be able to reproduce the observed trend in mass and [$\alpha$/Fe] ratios. However, they point out that a flatter IMF (skewed to high-mass stars) is needed to achieve the best agreement between observations and model. A slightly top-heavy IMF produces more massive stars, which in turn enrich the ISM more efficiently allowing for more metal-rich galaxies and a better agreement with observed abundances. \par

Similar high enhancements were found by \citet{puz06} in a large number of GCs in early-type galaxies. In their study they find significantly high [$\alpha$/Fe] abundances derived from Lick line index measurements, of the order of [$\alpha$/Fe] > 0.5 dex. They compare these ratios with supernova yield models and find that a significant contribution from stars with masses $>$20 M$_{\odot}$ are needed to reach [$\alpha$/Fe] ratios $\gtrsim$ 0.4 dex. In a comparable scenario to the one in NGC 5128, to explain the high [$\alpha$/Fe] ratios in the elliptical galaxies they proposed that in order to produce this amount of enriched material one needs to consider stellar populations composed of high-mass stars only, truncating the lower end of the IMF (< 20 M$_{\odot}$), in combination with a relatively high star-formation rate \citep{goo05,puz06}. We point out that these empirical results appear only in their early-type galaxy sample, and is absent in the spirals studied as part of their work, suggesting an intrinsic difference in the formation histories of different galaxy types. \par

As shown with yellow points in Figure \ref{fig:alpha}, recent studies have found higher metallicities, [Fe/H], and enhanced [$\alpha$/Fe] ratios for stars in the MW bulge \citep{mat90, ben13, jon14,gon15}, similar to what is observed for the [Mg/Fe] and [Ti/Fe] ratios in the metal-rich GCs in NGC 5128. In principle, this might suggest similar formation histories with rapid chemical enrichment for NGC 5128 and the MW bulge. Photometric observations of halo stars in NGC 5128 analysed by \citet{rej11} appear to support a rapid chemical enrichment scenario. \citet{rej11} find that their CMD observations agree with older populations with high metallicities and enhanced $\alpha$ abundances. \par

We point out that previous abundance studies of GCs in NGC 5128 using line indices and low-resolution spectroscopic observations have not always found higher [$\alpha$/Fe] ratios than in the MW. \citet{bea08} and \citet{woo10} estimated that the mean [$\alpha$/Fe] ratios of the GC system in NGC 5128 are lower than the mean ratios measured in the MW. However, this conclusion is not unexpected given the large systematic uncertainties accompanying line index studies \citep{bro06}. 

By visually inspecting Figure \ref{fig:alpha}, it appears that the [Mg/Fe] ratios throughout the whole metallicity range are more enhanced than [Ca/Fe] and [Ti/Fe]. As pointed out above, this behaviour has been observed in previous studies of dSph galaxies. Assuming this slight enhancement seen in [Mg/Fe] is not artificial, one possible explanation for differences between [(Ca, Ti)/Fe]  and [Mg/Fe] is the lack of hypernovae in NGC 5128. According to \citet{nak01} the $\alpha$-process takes place in hypernovae, creating $^{44}$Ca and $^{48}$Ti. Therefore an environment with a lower number of hypernovae would appear to have lower [Ca/Fe] and [Ti/Fe] ratios, compared to [Mg/Fe]. This was the explanation suggested by \citet{ven04} for the difference in abundance ratios for [(Ca, Ti)/Fe]  and [Mg/Fe] in dwarf spheroidals. \par

 \subsection{Light elements: Na}
 We are able to measure [Na/Fe] abundance ratios for 9 out of the 20 clusters studied in this work. Light elements in general are of special interest given that recent studies have shown that these might not be entirely monometallic in GC populations, compared to [Fe/H] and other $\alpha$-elements. We further discuss the observed star-to-star variations in GCs in Section \ref{MP}. \par
\begin{figure}
            {\includegraphics[scale=0.5]{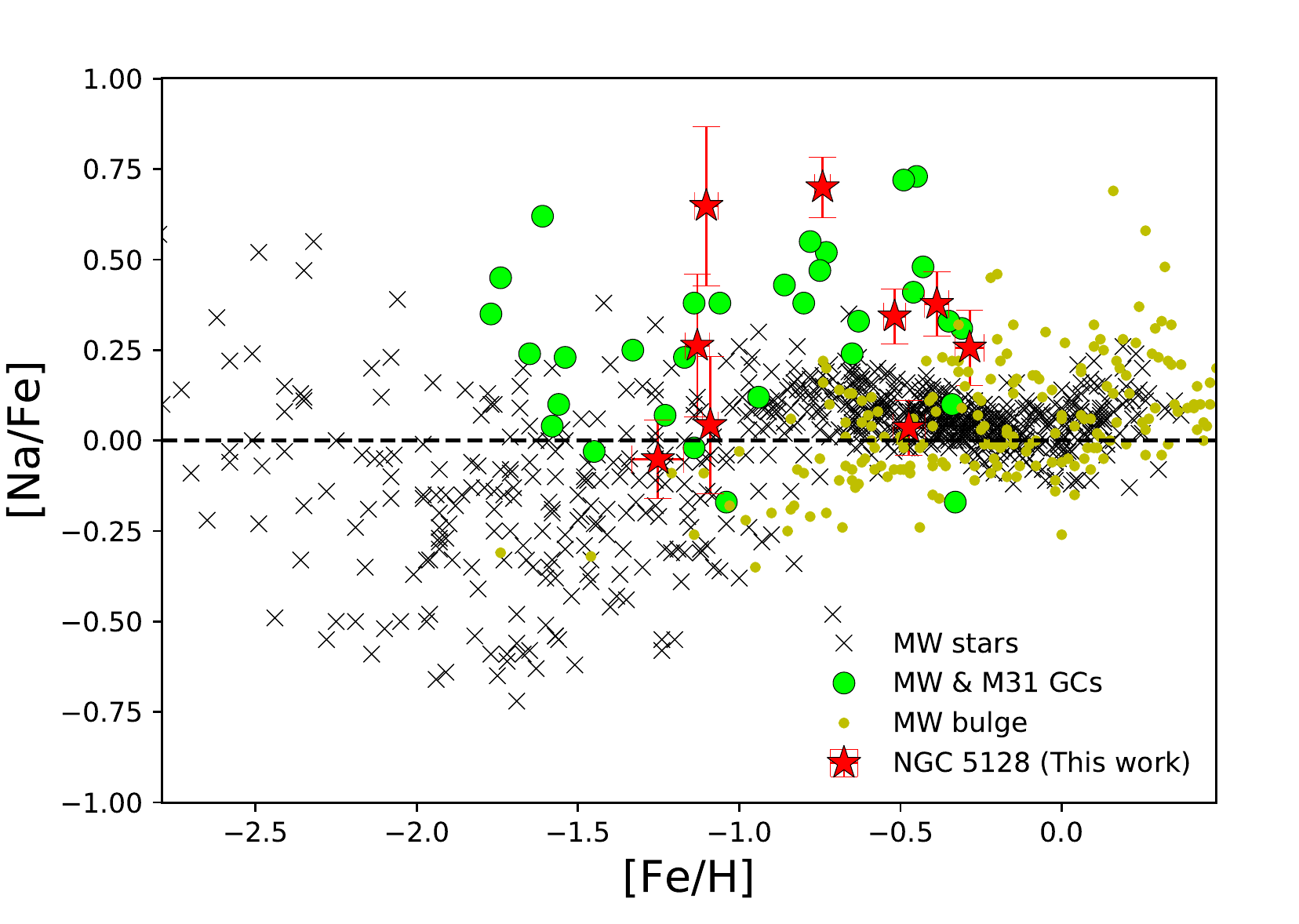}}
      \caption{[Na/Fe] abundance ratios as a function of [Fe/H]. Symbols similar to Figure \ref{fig:alpha}, with the exception of green circles here are integrated-light measurements of GCs in the MW and M31 by \citet{col14} and \citet{col16}.}
         \label{fig:Naabun}
   \end{figure}

We measured Na abundances using two wavelength bins, 5670-5700 \r{A} and 6148-6168 \r{A} covering the Na doublets 5682/5688 \r{A} and 6154/6160 \r{A}, respectively. This element is particularly difficult to measure as these lines are relatively weak. Given the quality of the spectra in the wavelength range of 5670-5700 \r{A}, the strongest of the two doublets, we are only able to measure Na abundances for 9 clusters. We note, however, that measurements for both bins are only available for 7 out of these 9 clusters. Whenever Na abundances are available for both bins, 5670-5700 \r{A} and 6148-6168 \r{A}, we find consistent Na measurements between the bins. In Figure \ref{fig:Naabun} we show our measured [Na/Fe] ratios as a function of metallicity. We see clearly elevated [Na/Fe] ratios in at least 5 GCs in NGC 5128, with higher values than those measured in field stars in the MW. Of course we currently do not have information about Na abundances of individual stars in NGC 5128; however, similar enhancements have been measured in integrated-light spectra of GCs in the MW and M 31, shown as green circles in Figure \ref{fig:Naabun} \citep{col14,col16}. \par
Clear correlations have been observed in Na and O abundances in MW GCs and extragalactic ones. We look for possible correlations between Na and any of the $\alpha$- and Fe-peak elements studied here, but find no such behaviour. \par
 
 \subsection{Fe-peak elements}
In this section we present our measurements for the Fe-peak elements: Cr, Mn, and Ni. In general, Fe-peak elements are suggested to be produced in SN Ia, however it is still not clear if all of the elements form in the same nucleosynthetic process. In spite of that, Fe-peak elements are of interest since they track the production of Fe \citep{iwa99}.  \par

\begin{figure*}
            {\includegraphics[scale=0.7]{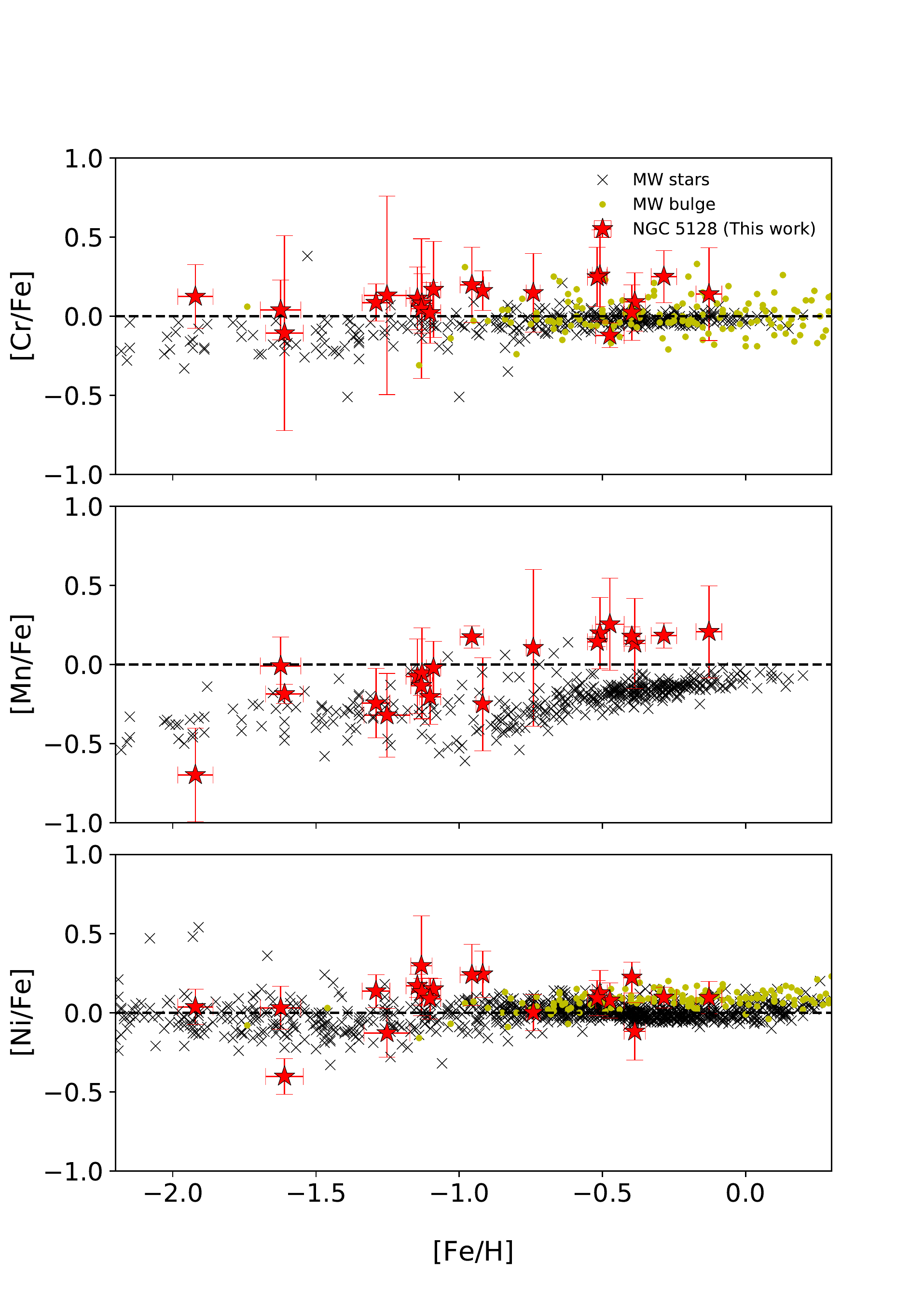}}
      \caption{Top: [Cr/Fe] vs. [Fe/H]. Middle: [Mn/Fe] vs. [Fe/H]. Bottom:  [Ni/Fe] vs. [Fe/H]. Symbols as in Figure \ref{fig:alpha}.}
         \label{fig:fepeak}
   \end{figure*}

Studies of Galactic stars show that the [Cr/Fe] abundance ratios are very close to [Cr/Fe]$\sim$ 0 throughout most of the metallicity coverage, as seen in the top panel of Figure \ref{fig:fepeak}. For NGC 5128 we also see that the measured [Cr/Fe] are close to 0, within the errors of the individual measurements. However, we see an overall trend where the [Cr/Fe] values for NGC 5128 are slightly above solar. We measure a weighted mean of [Cr/Fe] = $+$0.06 $\pm$ 0.02 for the full GC sample in NGC 5128, slightly higher than the mean ratio observed in MW field stars, [Cr/Fe] = $-$0.05 $\pm$ 0.01. The mean [Cr/Fe] seen in NGC 5128 is more comparable to that measured for the MW bulge stars, [Cr/Fe] = $+$0.01 $\pm$ 0.01 (Figure \ref{fig:fepeak} yellow points), and certainly found within the spread of measured [Cr/Fe] in the Galactic bulge.  \par

Although most of the Fe-peak elements behave somewhat similarly with Fe ratios close to 0, it is well known that Mn is depleted by $\sim-$ 0.5 dex in some metal-poor MW halo stars and GCs \citep{gra89, sob06, rom11}, including extragalactic GCs (see middle panel in Figure \ref{fig:fepeak}). The MW [Mn/Fe] ratios appear to increase towards solar values at higher metallicities. This behaviour has not been well understood, especially since the actual formation of Mn continues to be uncertain \citep{tim95, she03}. However, a possible explanation for this trend is that Mn is underproduced in SN II but overproduced in SN Ia, or a more accepted scenario is that the amount of Mn created is dependent on the metallicity of the progenitor \citep{gra89}. Given that several studies find that this trend in [Mn/Fe] could be caused by NLTE effects \citep{bat15, ber08}, we then make relative comparisons of our measured [Mn/Fe] to studies assuming LTE. It appears that once the NLTE corrections are applied, the trend disappears and follows a similar behaviour as the other Fe-peak elements. \par
Regarding our abundance measurements for NGC 5128 from Figure \ref{fig:fepeak} we can visually identify a trend of low [Mn/Fe] ratios at low metallicities, increasing towards solar, and above solar, at higher [Fe/H]. Although this is in agreement with similar trends observed in MW Galactic star and GC LTE studies, we see a slight offset in the trend of GCs in NGC 5128 compared to that of Galactic stars. For metallicities [Fe/H] > $-$1.0 dex, the [Mn/Fe] seem to reach higher abundances than the average MW field star population. \par
Ni also shows a relatively flat trend, with [Ni/Fe] abundance ratios close to 0, as shown in red stars in the bottom panel of Figure \ref{fig:fepeak}. The overall trend seen in our [Ni/Fe] ratios for NGC 5128 is consistent with that of studies in the MW. Taking the weighted mean values of the full GC sample in NGC 5128, we measure [Ni/Fe] = $+$0.09 $\pm$ 0.04, compared to a slightly lower mean [Ni/Fe] ratio of $-$0.02 $\pm$ 0.01 for MW field stars.\par
Overall, we find that our Fe-peak abundances follow similar trends as those seen in MW field stars, however, we identify a slight enhancement in the Fe-peak measurements from the GC system in NGC 5128. \par

 \subsection{Multiple populations in GCs}\label{MP}
Studies have shown that the [Mg/Fe] abundance ratios of field stars in the MW, LMC and M 31 behave similarly to other $\alpha$-elements \citep{ben05,col09,gon11}. However, in the last decades in-depth research on individual stellar abundances in GCs has presented results that suggest the presence of intra-cluster Mg variations with respect to $\alpha$-elements in several Galactic clusters \citep{she96, kra97, gra04}. Furthermore, integrated-light studies of extragalactic GCs appear to also detect these same variations in Mg as lower [Mg/Fe] when compared to other [$\alpha$/Fe] ratios. Through a study of the integrated-light of 31 GCs in M 31 \citet{col14} finds evidence for intra-cluster Mg variations. More specifically, they find that the more metal-poor, [Fe/H] < $-0.7$, GCs in M 31 are biased to low [Mg/Fe] ratios, hinting at a depletion of Mg. They measure [Mg/Fe] $\lesssim$ 0.0 for GCs with metallicities [Fe/H] < $-$1.5. Similarly, studying the integrated light of extragalactic GCs in Fornax, \citet{lar12} measure lower [Mg/Fe] than [Ca/Fe] and [Ti/Fe]. From Figure \ref{fig:alpha} we clearly see that even in the low metallicity regime the [Mg/Fe] values are comparable to, or even higher than, those measured for [Ca/Fe] and [Ti/Fe]. Compared to the observations of \citet{col14} regarding the metal-poor GCs in M 31, we instead measure [Mg/Fe] ratios above solar abundances. These values suggest an absence of intra-cluster Mg variations in the GC populations in NGC 5128. \par
Similar to Mg, Na has exhibited star-to-star abundance variations in the GC population in the MW \citep[for a recent review see ][]{gra04}. In the last years studies have found indirect evidence for abundance variations in extragalactic GCs (M 31, LMC, WLM, IKN) that support the idea of intra-cluster abundance variations in Na \citep{col09, muc09,lar14}. Specifically for Na, these intra-cluster variations are perceived as significantly elevated [Na/Fe] in integrated-light analysis of both galactic and extragalactic GCs. We find elevated [Na/Fe] abundance ratios for 5 GCs in NGC 5128, shown as red stars in Figure ~\ref{fig:Naabun}. The enhancement in these GCs is higher than any abundances previously measured in any MW field stars, halo, disk or bulge. From the context of multiple populations in GCs, these enhancement in the Na abundances can in principle show evidence of star-to-star variations. \par


\section{Conclusions}\label{con}
Through the analysis of integrated-light observations taken with the X-Shooter spectrograph on ESO's VLT and synthetic spectra created based on theoretical $\alpha$-enhanced isochrones by the Dartmouth group, we obtain accurate metallicities for 20 GCs. The GCs studied here have metallicities ranging between [Fe/H] = $-$1.92 and $-$0.13 dex. We measure $\alpha$-elements (Mg, Ca, and Ti), as well as a single light element (Na), and Fe-peak elements (Cr, Mn, and Ni). The new results of our work can be summarised as follows:

   \begin{enumerate}
      \item Comparing our inferred metallicities to measurements from \citet{bea08}, we find excellent agreement between both works. 
   \item When comparing the four GC [Ca/Fe] abundance measurements to those by \citet{col13} we find that three out of four clusters are in good agreement, within 1$\sigma$. The [Ca/Fe] measurement of the fourth GC in \citet{col13} appears to be a 3-$\sigma$ outlier compared to our inferred abundances.
   \item We find higher enhancement ($\sim$ 0.1 dex) in $\alpha$-elements for the metal-poor ([Fe/H$<-$0.75 dex) GCs in NGC 5128, than what is observed in the MW. This could hint at a top-heavy IMF, a scenario supported by semi-analytical models of galactic chemical evolution of early-type galaxies. \citet{arr10} find that a slightly top-heavy IMF is essential to match the observations with models. However, we recommend to extend the GC sample size to firmly confirm such a scenario.
   \item We find [Mg/Fe] ratios to be slightly higher than the [Ca/Fe] and [Ti/Fe] ratios. We suggest a lack of hypernovae in NGC 5128 as a possible explanation for this difference. 
   \item In the context of multiple populations of GCs in NGC 5128, we measure Na in 9 clusters. We find enhanced [Na/Fe] ratios for 5 GCs in our sample, providing evidence of star-to-star variations in these GCs. We find an absence of intra-cluster Mg variations in all 20 clusters. 
   \item We obtain the first measurements of Cr, Mn and Ni for 20 GCs in NGC 5128, and find that the overall abundances follow similar trends as those seen in the MW field stars. We see a slight enhancement ($<$ 0.1 dex) in the mean Fe-peak abundances, when compared to those seen in the MW. 
    \end{enumerate}	
    
We point out that the work presented here is based on GC observations with exposure times of $\lesssim$ 30 minutes. Longer exposure times would provide better S/N on all three X-Shooter arms, increasing the integrated-light analysis potential especially for the NIR wavelengths. Such a coverage would allow for an in-depth study of the Al lines (i.e. 13123.38 and 13150.71 \r{A}) providing information on possible [Al/Fe] enhancements similar to those observed in Galactic GCs, contributing to our discussion of multiple populations in extragalactic GCs. \par
The general agreement between existing metallicities and Ca abundance measurements and those presented as part of this work is extremely encouraging for future integrated-light analysis, reassuring that reliable metallicities and detailed chemical abundances can be obtained using intermediate-resolution observations.

\section*{Acknowledgements}
We thank A. Gonneau, Y.-P. Chen and M. Dries for their help and guidance during the X-Shooter reduction process. This research has made use of the NASA/IPAC Extragalactic Database (NED), which is operated by the Jet Propulsion Laboratory, California Institute of Technology, under contract with the National Aeronautics and Space Administration.








\appendix
\clearpage

\section{Abundances as a function of wavelength}\label{sec:A}
\begin{figure}
            {\includegraphics[scale=0.62]{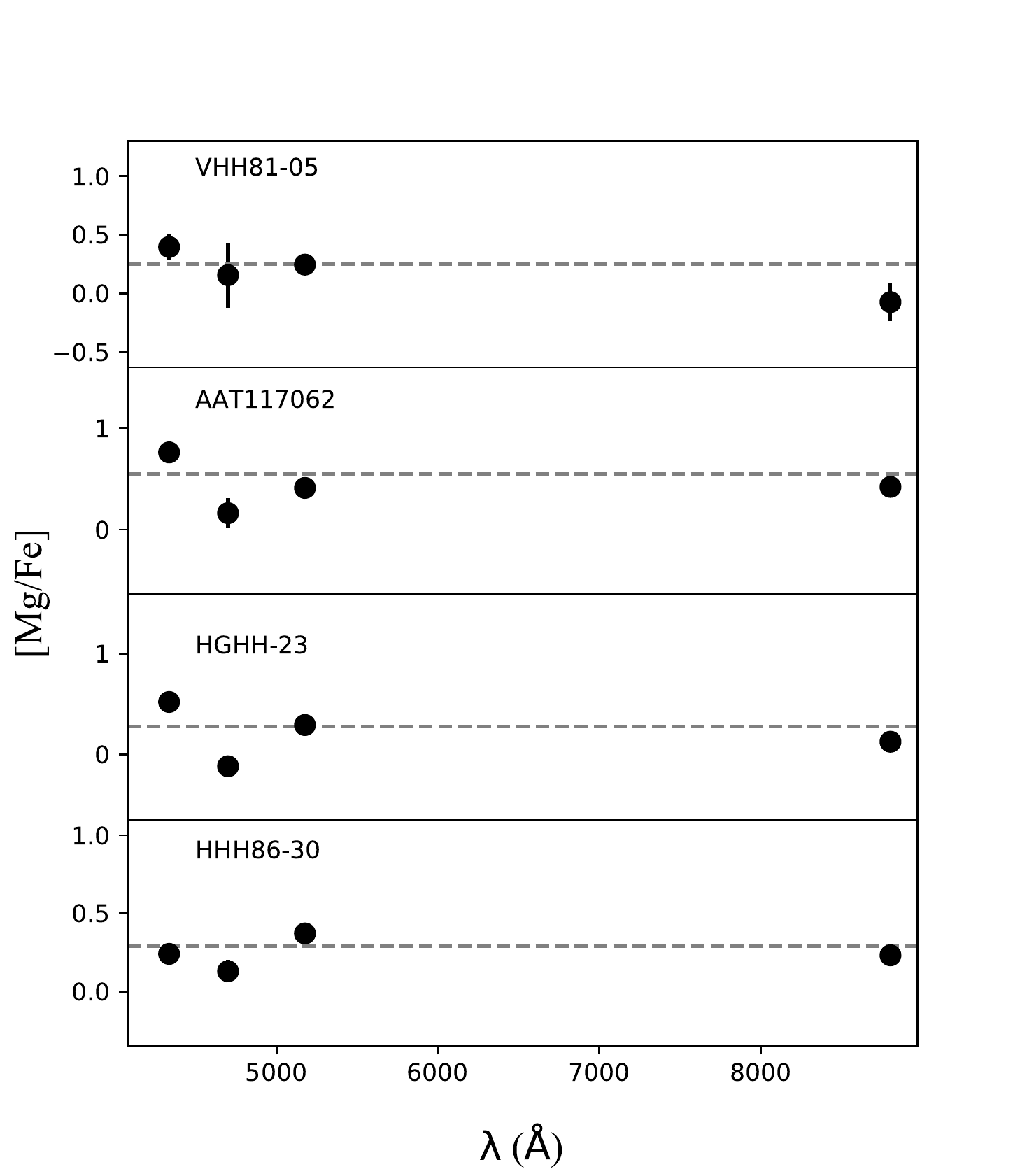}}
      \caption{[Mg/Fe] abundances as a function of wavelength for a selected sample of GCs in NGC 5128.}
         \label{fig:mg_app}
   \end{figure}

\begin{figure}
            {\includegraphics[scale=0.62]{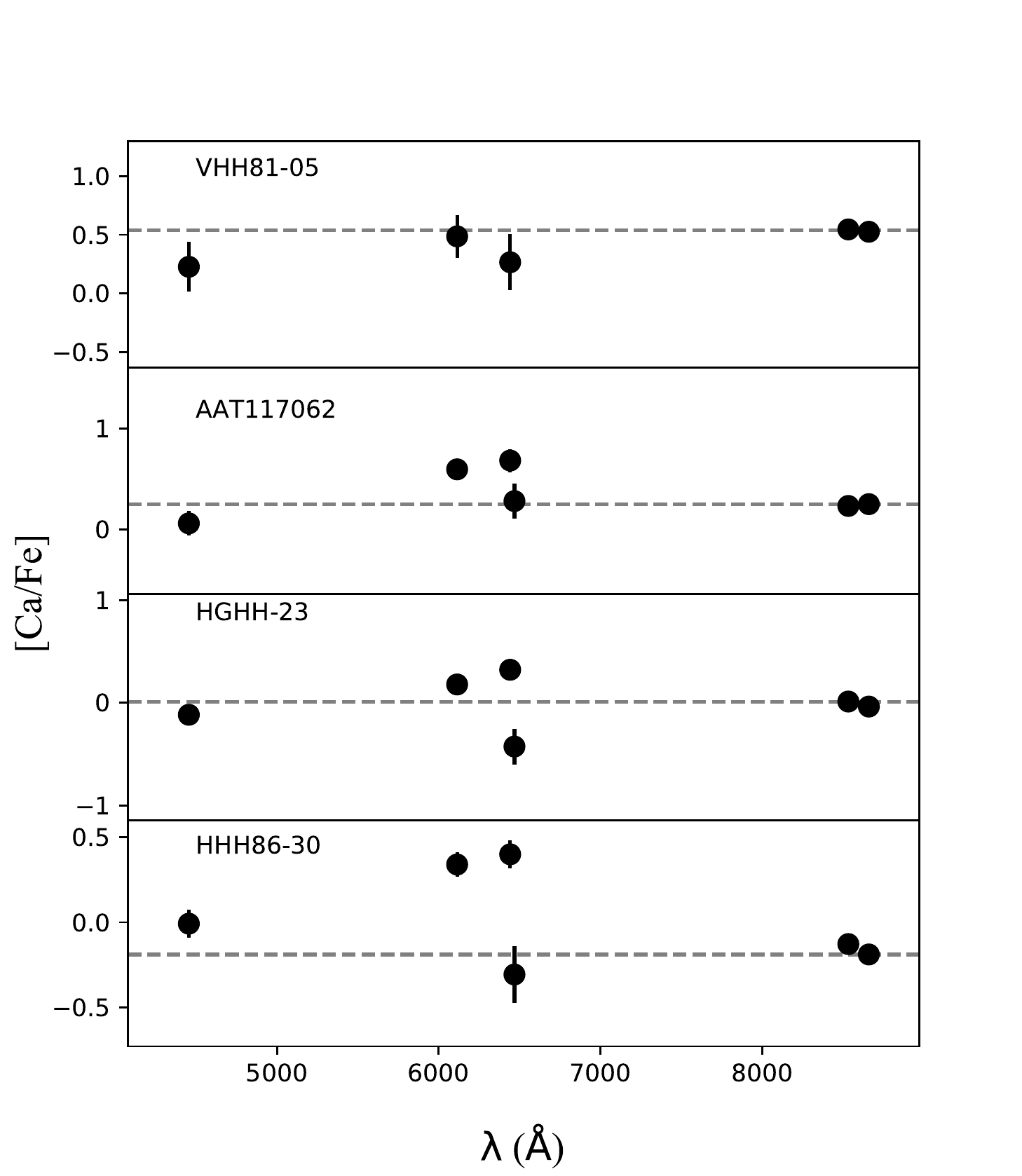}}
      \caption{[Ca/Fe] abundances as a function of wavelength for a selected sample of GCs in NGC 5128.}
         \label{fig:ca_app}
   \end{figure}

\begin{figure}
            {\includegraphics[scale=0.62]{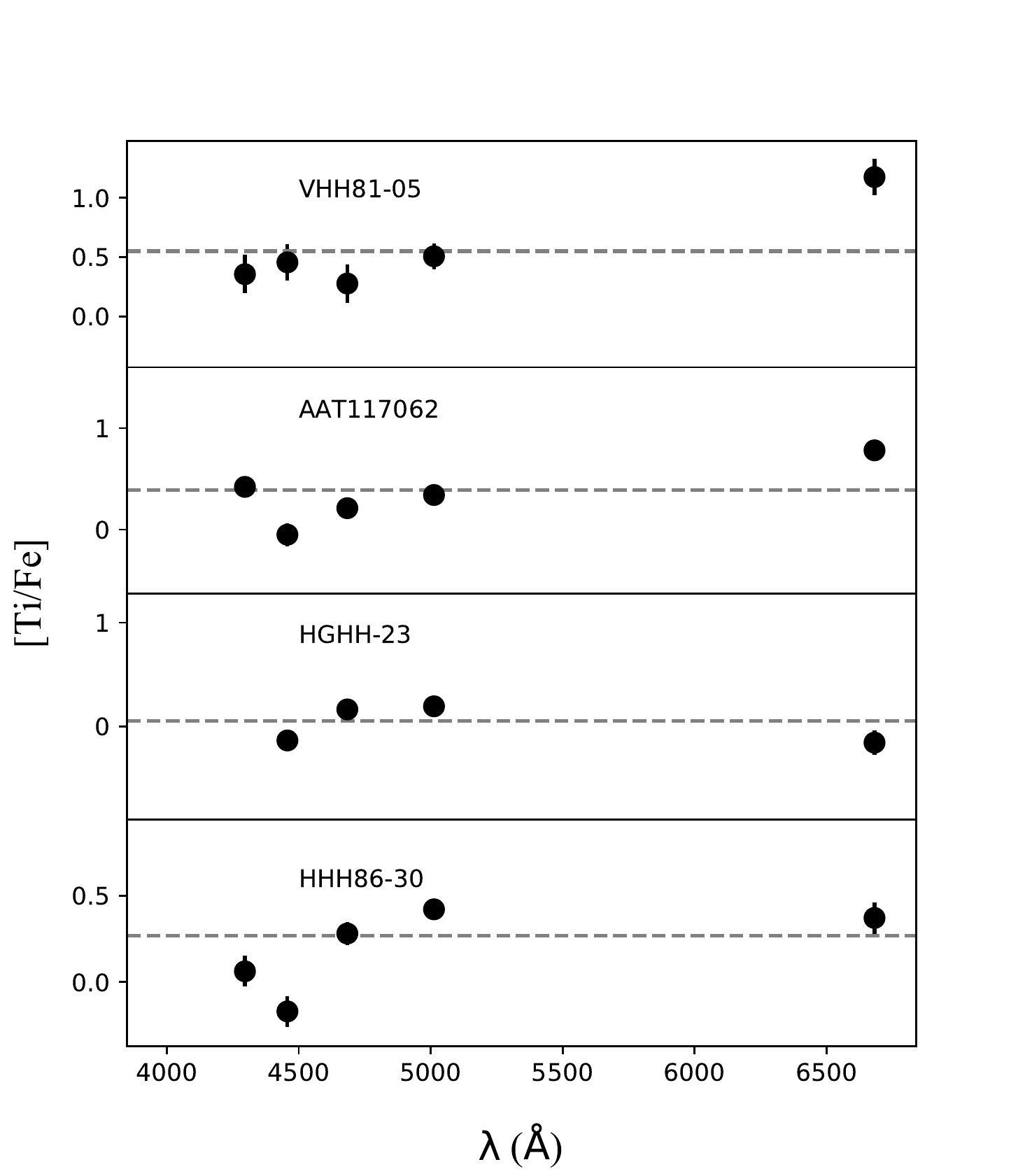}}
      \caption{[Ti/Fe] abundances as a function of wavelength for a selected sample of GCs in NGC 5128.}
         \label{fig:ti_app}
   \end{figure}

\begin{figure}
            {\includegraphics[scale=0.62]{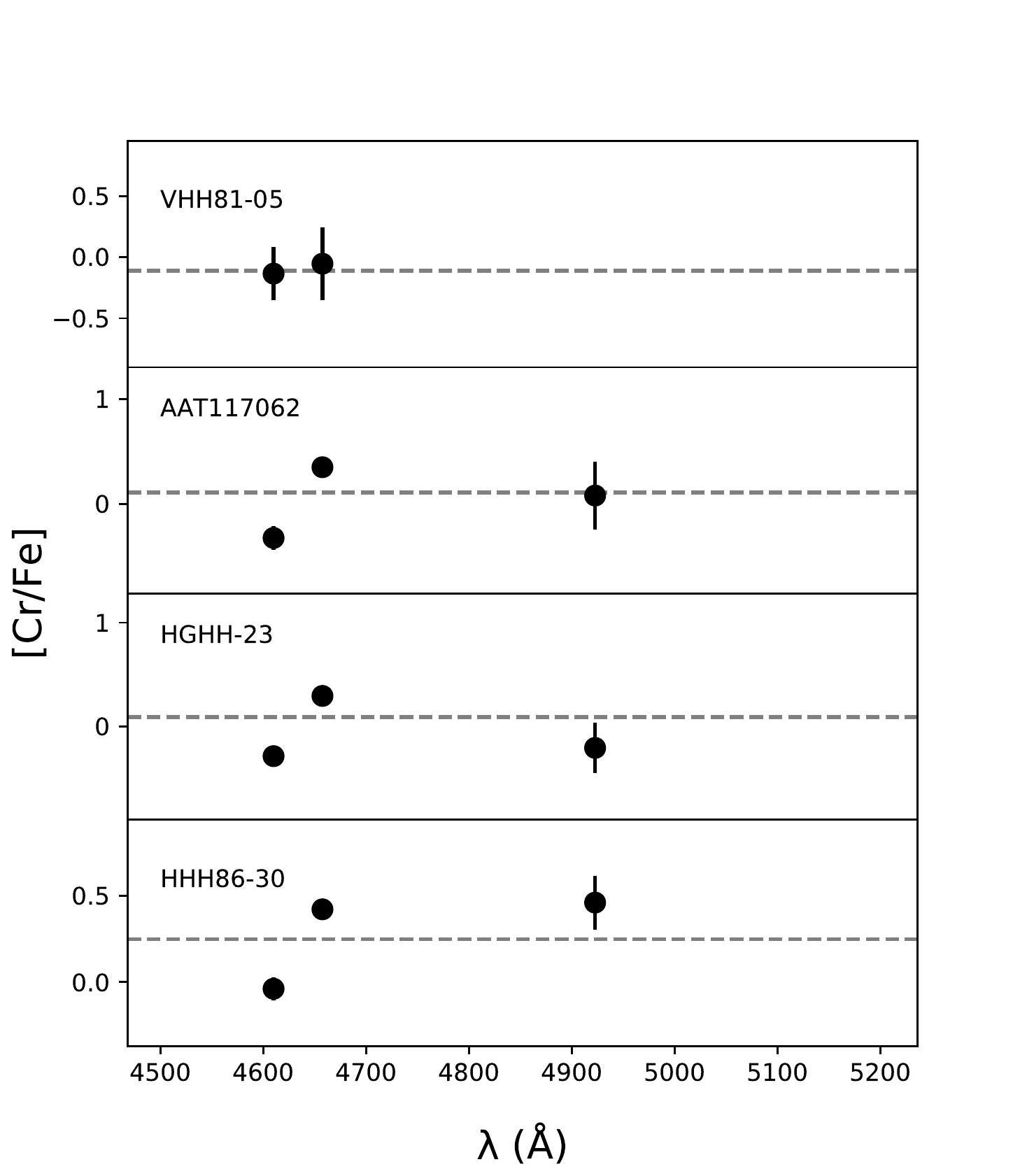}}
      \caption{[Cr/Fe] abundances as a function of wavelength for a selected sample of GCs in NGC 5128.}
         \label{fig:cr_app}
   \end{figure}

\begin{figure}
            {\includegraphics[scale=0.62]{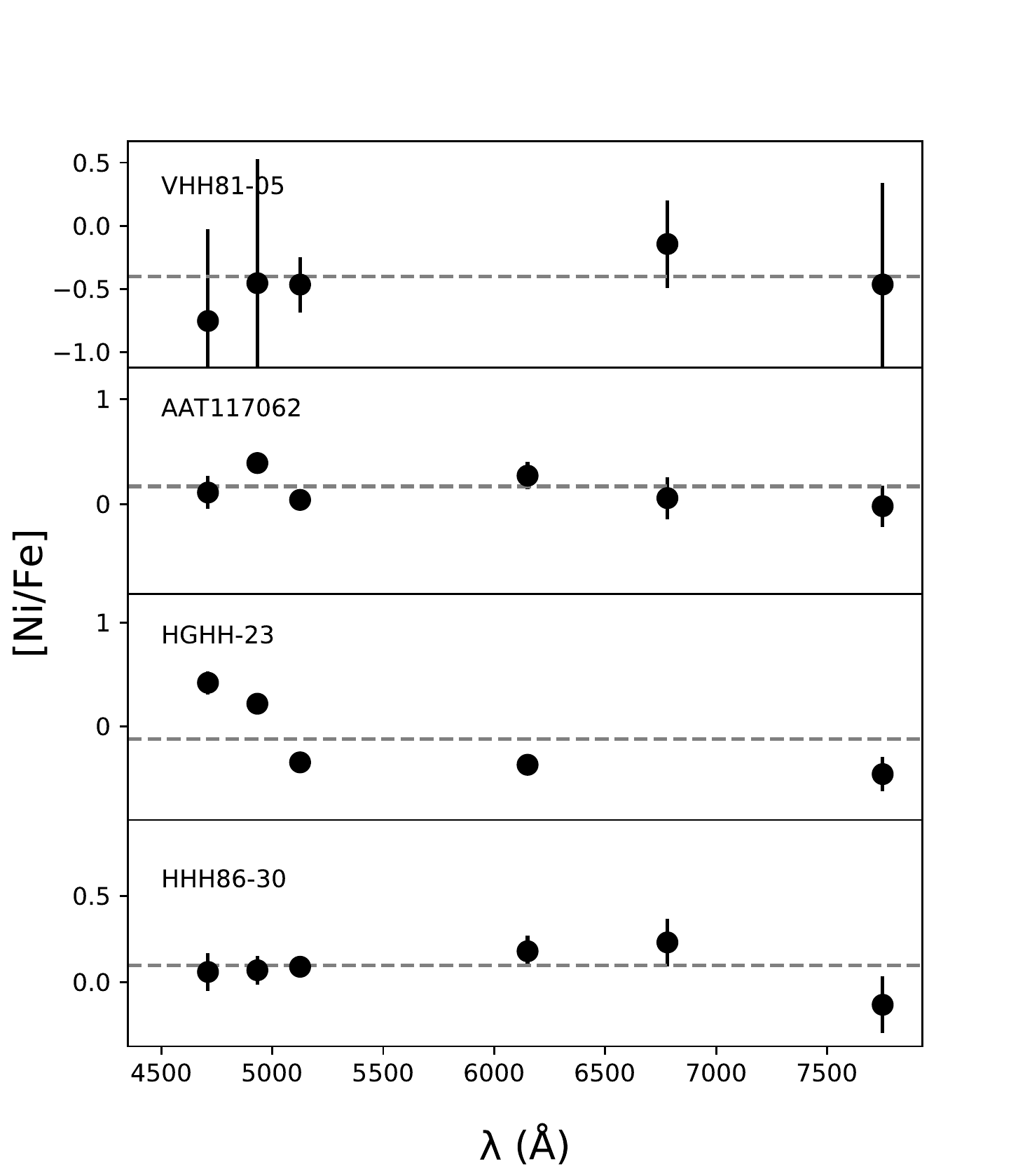}}
      \caption{[Ni/Fe] abundances as a function of wavelength for a selected sample of GCs in NGC 5128.}
         \label{fig:ni_app}
   \end{figure}

We present tables displaying the individual bin measurements for each of the GCs studied in this work. Additionally, to visually appreciate the dispersion between the individual measurements for a corresponding element with multiple bins, in Figures \ref{fig:mg_app} to \ref{fig:ni_app} we plot bin abundances as a function of wavelength for a selected sample of GCs. 

\begin{table}
\caption{Chemical abundances AAT117062}
\label{tab:1}
 \centering 
\begin{tabular}{cccc} 
 \hline  \hline \
Element& $Wavelength\, [$\AA$]$&Abundance &Error\\
  \hline\
{[Fe/H]}& 4400.0-4600.0& $-$1.265 & 0.021   \\
& 4600.0-4800.0& $-$1.194 & 0.033\\
& 4800.0-5000.0& $-$1.124 & 0.021 \\
& 5000.0-5200.0&  $-$1.184 &0.031 \\
& 6100.0-6300.0& $-$1.166 &0.041\\
& 6300.0-6500.0&  $-$1.154 &0.042 \\
& 6500.0-6700.0&  $-$1.225 &0.048 \\
& 6700.0-6800.0&  $-$0.974 &0.112 \\
&7400.0-7550.0 & $-$1.166 &0.061 \\
&8500.0-8700.0 &$-$0.834 &0.031 \\
& 8700.0-8850.0&  $-$0.984 &0.091  \\

{[Mg/Fe]}&4300.0-4370.0& $+$0.761 &0.054 \\
& 4690.0-4710.0 & $+$0.162 &0.146 \\
&5150.0-5200.0& $+$0.411&0.058\\
&8777.0-8832.0& $+$0.421 &0.066\\

{[Ca/Fe]}&4445.0-4465.0&  $+$0.062 &0.120  \\
&6100.0-6128.0&$+$0.595&0.083\\
&6430.0-6454.0 &$+$0.682 &0.113\\
&6459.0-6478.0 &$+$0.282 &0.175 \\
&8480.0-8586.0& $+$0.234 &0.037 \\
&8623.0-8697.0 &$+$0.252&0.038 \\

{[Ti/Fe]}&4270.0-4322.0 & $+$0.421 &0.088 \\
& 4440.0-4474.0 &$-$0.050 &0.115 \\
&4650.0-4718.0& $+$0.211 &0.090\\
&4980.0-5045.0& $+$0.341 &0.062 \\
&6584.0-6780.0 &$+$0.781 &0.080\\

{[Na/Fe]}&5670.0-5700.0& - & -\\
& 6148.0-6168.0 & - & -\\

{[Cr/Fe]}&4580.0-4640.0& $-$0.322 &0.112\\
&4640.0-4675.0& $+$0.350 &0.087\\
&4915.0-4930.0& $+$0.081 &0.323 \\

{[Mn/Fe]}& 4450.0-4515.0 &$-$0.438 &0.205 \\
&4750.0-4770.0 & $-$0.030 &0.115 \\

{[Ni/Fe]}& 4700.0-4720.0& $+$0.111 &0.157\\
&4910.0-4955.0&  $+$0.391 &0.091\\
&5075.0-5175.0 &  $+$0.041 &0.078\\
&6100.0-6200.0 & $+$0.271&0.130\\
&6760.0-6800.0 & $+$0.058 &0.201 \\
&7700.0-7800.0 &$-$0.018 &0.195\\
 \hline 
 \end{tabular}
\end{table}

\begin{table}
\caption{Chemical abundances HGHH-04}
\label{tab:2}
 \centering 
\begin{tabular}{cccc} 
 \hline  \hline \
Element& $Wavelength\, [$\AA$]$&Abundance &Error\\
  \hline\
{[Fe/H]}& 4400.0-4600.0&  $-$2.004 &0.082  \\
& 4600.0-4800.0& $-$1.824 &0.092 \\
& 4800.0-5000.0& $-$1.904 &0.062 \\
& 5000.0-5200.0&  $-$1.945 &0.061 \\
& 6100.0-6300.0& $-$1.535 &0.091\\
& 6300.0-6500.0&  $-$2.014 &0.112 \\
& 6500.0-6700.0&  $-$1.963 &0.192 \\
& 6700.0-6800.0&  $-$1.834 &0.322 \\
&7400.0-7550.0 & $-$2.184 &0.212 \\
&8500.0-8700.0 & $-$2.323 &0.112 \\
& 8700.0-8850.0&  $-$1.865 &0.181  \\

{[Mg/Fe]}&4300.0-4370.0& - & -\\
& 4690.0-4710.0 &$+$0.116 &0.355 \\
&5150.0-5200.0&$+$0.467 &0.083\\
&8777.0-8832.0&$-$0.093 &0.337\\

{[Ca/Fe]}&4445.0-4465.0&  $-$0.083 &0.385  \\
&6100.0-6128.0& $+$0.778 &0.219\\
&6430.0-6454.0 &$+$0.778&0.229\\
&6459.0-6478.0 &$+$0.046 &0.572 \\
&8480.0-8586.0& $+$0.527 &0.057 \\
&8623.0-8697.0 &$+$0.507 &0.064 \\

{[Ti/Fe]}&4270.0-4322.0 & $+$0.608 &0.191 \\
& 4440.0-4474.0 & $+$0.707 &0.21 \\
&4650.0-4718.0& $+$0.367 &0.238\\
&4980.0-5045.0& $+$0.437 &0.143 \\
&6584.0-6780.0 & $+$0.267 &1.791\\

{[Na/Fe]}&5670.0-5700.0& - & -\\
& 6148.0-6168.0 & - & -\\

{[Cr/Fe]}&4580.0-4640.0& $+$0.098 &0.278\\
&4640.0-4675.0& $+$0.118 &0.376\\
&4915.0-4930.0& $+$0.687 &1.212 \\

{[Mn/Fe]}& 4450.0-4515.0 & $-$0.153 &2.173 \\
&4750.0-4770.0 & $-$0.732&0.545 \\

{[Ni/Fe]}& 4700.0-4720.0& $+$0.297 &0.356\\
&4910.0-4955.0&  $+$0.067 &0.844\\
&5075.0-5175.0 &  $-$0.192 &0.278\\
&6100.0-6200.0 & $-$0.004 &0.355\\
&6760.0-6800.0 & $-$0.223 &0.605 \\
&7700.0-7800.0 & $+$0.328 &0.386\\
 \hline 
 \end{tabular}
\end{table}

\begin{table}
\caption{Chemical abundances HGHH-06}
\label{tab:3}
 \centering 
\begin{tabular}{cccc} 
 \hline  \hline \
Element& $Wavelength\, [$\AA$]$&Abundance &Error\\
  \hline\
{[Fe/H]}& 4400.0-4600.0&  $-$0.955 &0.041  \\
& 4600.0-4800.0& $-$0.855 &0.063 \\
& 4800.0-5000.0& $-$0.876 &0.038 \\
& 5000.0-5200.0&  $-$0.995 &0.061 \\
& 6100.0-6300.0&  $-$0.886 &0.076\\
& 6300.0-6500.0&  $-$0.925 &0.090 \\
& 6500.0-6700.0&  $-$0.955 &0.109 \\
& 6700.0-6800.0&  $-$1.055 &0.241 \\
&7400.0-7550.0 & $-$0.835 &0.090 \\
&8500.0-8700.0 & $-$0.996 &0.067 \\
& 8700.0-8850.0&  $-$0.824 &0.136  \\

{[Mg/Fe]}&4300.0-4370.0& $+$0.792 &0.092 \\
& 4690.0-4710.0 & $+$0.523 &0.166 \\
&5150.0-5200.0& $+$0.632 &0.054\\
&8777.0-8832.0& $+$0.584 &0.084\\

{[Ca/Fe]}&4445.0-4465.0&  $-$0.077 &0.211  \\
&6100.0-6128.0& $+$0.857 &0.096\\
&6430.0-6454.0 & $+$0.793 &0.134\\
&6459.0-6478.0 & $+$0.054 &0.293 \\
&8480.0-8586.0& $+$0.346 &0.025 \\
&8623.0-8697.0 & $+$0.304 &0.024 \\

{[Ti/Fe]}&4270.0-4322.0 & $+$0.063 &0.178 \\
& 4440.0-4474.0 & $+$0.084 &0.172 \\
&4650.0-4718.0& $+$0.483 &0.122\\
&4980.0-5045.0& $+$0.293 &0.074 \\
&6584.0-6780.0 & $+$0.503 &0.114\\

{[Na/Fe]}&5670.0-5700.0& - & -\\
& 6148.0-6168.0 & - & - \\

{[Cr/Fe]}&4580.0-4640.0& $+$0.012 &0.141\\
&4640.0-4675.0& $+$0.292 &0.144\\
&4915.0-4930.0& $+$0.433 &0.492 \\

{[Mn/Fe]}& 4450.0-4515.0 & $-$0.736 &0.451 \\
&4750.0-4770.0 & $-$0.149 &0.208 \\

{[Ni/Fe]}& 4700.0-4720.0& $-$0.003 &0.269\\
&4910.0-4955.0& $+$0.593 &0.147\\
&5075.0-5175.0 & $+$0.243 &0.103\\
&6100.0-6200.0 & $+$0.123 &0.192\\
&6760.0-6800.0 & $-$0.487 &0.281 \\
&7700.0-7800.0 & $+$0.214 &0.282\\
 \hline 
 \end{tabular}
\end{table}

\begin{table}
\caption{Chemical abundances HGHH-07}
\label{tab:4}
 \centering 
\begin{tabular}{cccc} 
 \hline  \hline \
Element& $Wavelength\, [$\AA$]$&Abundance &Error\\
  \hline\
{[Fe/H]}& 4400.0-4600.0&  $-$1.284 &0.025  \\
& 4600.0-4800.0& $-$1.112 &0.033 \\
& 4800.0-5000.0& $-$1.086 &0.023 \\
& 5000.0-5200.0&  $-$1.134 &0.026 \\
& 6100.0-6300.0& $-$1.013 &0.043\\
& 6300.0-6500.0&  $-$0.929 &0.050 \\
& 6500.0-6700.0&  $-$1.055 &0.078 \\
& 6700.0-6800.0&  $-$1.244 &0.172 \\
&7400.0-7550.0 & $-$1.125 &0.076 \\
&8500.0-8700.0 & $-$1.196 &0.046 \\
& 8700.0-8850.0&  $-$0.886 &0.090  \\

{[Mg/Fe]}&4300.0-4370.0& $+$0.924 &0.064 \\
& 4690.0-4710.0 & $+$0.456 & 0.116 \\
&5150.0-5200.0& $+$0.486 &0.041\\
&8777.0-8832.0& $+$0.376 &0.070\\

{[Ca/Fe]}&4445.0-4465.0&  $+$0.116 &0.116  \\
&6100.0-6128.0& $+$0.807 &0.077\\
&6430.0-6454.0 & $+$0.596 &0.125\\
&6459.0-6478.0 & $-$0.005 &0.219 \\
&8480.0-8586.0& $+$0.505 &0.036 \\
&8623.0-8697.0 & $+$0.436 &0.036 \\

{[Ti/Fe]}&4270.0-4322.0 & $+$0.104 &0.133 \\
& 4440.0-4474.0 & $+$0.056 &0.120 \\
&4650.0-4718.0& $+$0.484 &0.087\\
&4980.0-5045.0& $+$0.215 &0.070 \\
&6584.0-6780.0 & $+$0.414 &0.109\\

{[Na/Fe]}&5670.0-5700.0& $+$0.145 &0.193 \\
& 6148.0-6168.0 & $+$0.534 &0.292 \\

{[Cr/Fe]}&4580.0-4640.0& $-$0.054&0.097\\
&4640.0-4675.0& $+$0.224 &0.129\\
&4915.0-4930.0& $+$0.585 &0.276 \\

{[Mn/Fe]}& 4450.0-4515.0 & $-$0.505 &0.252 \\
&4750.0-4770.0 & $+$0.065 &0.134 \\

{[Ni/Fe]}& 4700.0-4720.0& $+$0.044 &0.194\\
&4910.0-4955.0& $+$0.224 &0.134\\
&5075.0-5175.0 & $-$0.005 &0.095\\
&6100.0-6200.0 & $+$0.245 &0.146\\
&6760.0-6800.0 & $+$0.495 &0.193 \\
&7700.0-7800.0 & $+$0.047 &0.222\\
 \hline 
 \end{tabular}
\end{table}

\begin{table}
\caption{Chemical abundances HGHH-11}
\label{tab:5}
 \centering 
\begin{tabular}{cccc} 
 \hline  \hline \
Element& $Wavelength\, [$\AA$]$&Abundance &Error\\
  \hline\
{[Fe/H]}& 4400.0-4600.0&  -0.444 &0.012  \\
& 4600.0-4800.0& -0.126 &0.028 \\
& 4800.0-5000.0& -0.585 &0.021 \\
& 5000.0-5200.0&  -0.624 &0.031 \\
& 6100.0-6300.0& -0.574 &0.042\\
& 6300.0-6500.0&  -0.546 &0.047 \\
& 6500.0-6700.0&  -0.755 &0.079 \\
& 6700.0-6800.0&  -0.654 &0.092 \\
&7400.0-7550.0 & -0.656 &0.058 \\
&8500.0-8700.0 & -0.484 &0.042 \\
& 8700.0-8850.0&  -0.624 &0.07  \\

{[Mg/Fe]}&4300.0-4370.0& $+$0.400 &0.112 \\
& 4690.0-4710.0 & $+$0.081 &0.128 \\
&5150.0-5200.0& $+$0.468 &0.061\\
&8777.0-8832.0& $+$0.090 &0.069\\

{[Ca/Fe]}&4445.0-4465.0& $+$0.004&0.126  \\
&6100.0-6128.0& $+$0.103 &0.089\\
&6430.0-6454.0 & $-$0.020 &0.112\\
&6459.0-6478.0 & $-$0.649 &0.224 \\
&8480.0-8586.0& $-$0.076 &0.046 \\
&8623.0-8697.0 & $-$0.120 &0.047 \\

{[Ti/Fe]}&4270.0-4322.0 & $-$0.160 &0.158 \\
& 4440.0-4474.0 & $+$0.000 &0.130 \\
&4650.0-4718.0& $+$0.260 &0.098\\
&4980.0-5045.0&  $+$0.230 &0.084 \\
&6584.0-6780.0 & $-$0.511 &0.215\\

{[Na/Fe]}&5670.0-5700.0& $+$0.064 &0.184 \\
& 6148.0-6168.0 & $-$0.059 &0.326 \\

{[Cr/Fe]}&4580.0-4640.0& $-$0.241 &0.102\\
&4640.0-4675.0& $-$0.022 &0.094\\
&4915.0-4930.0& $-$0.151 &0.323 \\

{[Mn/Fe]}& 4450.0-4515.0 & $-$0.158 &0.180 \\
&4750.0-4770.0 & $+$0.419 &0.119 \\

{[Ni/Fe]}& 4700.0-4720.0& $+$0.159 &0.166\\
&4910.0-4955.0& $+$0.319 &0.119  \\
&5075.0-5175.0 &  $-$0.252 &0.118\\
&6100.0-6200.0 & $+$0.222 &0.122\\
&6760.0-6800.0 & $-$0.260 &0.206 \\
&7700.0-7800.0 & $+$0.145 &0.200\\
 \hline 
 \end{tabular}
\end{table}

\begin{table}
\caption{Chemical abundances HGHH-17}
\label{tab:6}
 \centering 
\begin{tabular}{cccc} 
 \hline  \hline \
Element& $Wavelength\, [$\AA$]$&Abundance &Error\\
  \hline\
{[Fe/H]}& 4400.0-4600.0& $-$1.194 &0.022  \\
& 4600.0-4800.0& $-$1.080 &0.035 \\
& 4800.0-5000.0& $-$1.055 &0.030 \\
& 5000.0-5200.0&  $-$1.025 &0.030 \\
& 6100.0-6300.0& $-$0.946 &0.044\\
& 6300.0-6500.0&  $-$1.124 &0.052 \\
& 6500.0-6700.0&  $-$1.114 &0.071 \\
& 6700.0-6800.0&  $-$0.974 &0.102 \\
&7400.0-7550.0 & $-$0.997 &0.069 \\
&8500.0-8700.0 & $-$1.126 &0.037 \\
& 8700.0-8850.0& $-$0.934 &0.066  \\

{[Mg/Fe]}&4300.0-4370.0& $+$0.755 &0.062 \\
& 4690.0-4710.0 & $+$0.276 &0.134 \\
&5150.0-5200.0& $+$0.485 &0.031\\
&8777.0-8832.0& $+$0.596 &0.066\\

{[Ca/Fe]}&4445.0-4465.0&  $+$0.316 &0.094  \\
&6100.0-6128.0& $+$0.492 &0.085\\
&6430.0-6454.0 & $+$0.745 &0.104\\
&6459.0-6478.0 & $+$0.167 &0.172 \\
&8480.0-8586.0& $+$0.455&0.026 \\
&8623.0-8697.0 & $+$0.427 &0.027 \\

{[Ti/Fe]}&4270.0-4322.0 & $+$0.215 &0.104 \\
& 4440.0-4474.0 & $+$0.126 &0.120 \\
&4650.0-4718.0& $+$0.426 &0.084\\
&4980.0-5045.0& $+$0.266 &0.056 \\
&6584.0-6780.0 & $+$0.226 &0.124\\

{[Na/Fe]}&5670.0-5700.0& $+$0.043 &0.189 \\
& 6148.0-6168.0 & - & -\\

{[Cr/Fe]}&4580.0-4640.0& $-$0.084 &0.105\\
&4640.0-4675.0&  $+$0.215 &0.098\\
&4915.0-4930.0& $+$0.935 &0.202 \\

{[Mn/Fe]}& 4450.0-4515.0 & $-$0.235 &0.175 \\
&4750.0-4770.0 & $+$0.095 &0.130 \\

{[Ni/Fe]}& 4700.0-4720.0& $+$0.174 &0.143\\
&4910.0-4955.0& $+$0.155 &0.122\\
&5075.0-5175.0 &  $-$0.004 &0.075\\
&6100.0-6200.0 & $+$0.384 &0.112\\
&6760.0-6800.0 & $+$0.387 &0.164 \\
&7700.0-7800.0 & $+$0.168 &0.196\\
 \hline 
 \end{tabular}
\end{table}

\begin{table}
\caption{Chemical abundances HGHH-21}
\label{tab:7}
 \centering 
\begin{tabular}{cccc} 
 \hline  \hline \
Element& $Wavelength\, [$\AA$]$&Abundance &Error\\
  \hline\
{[Fe/H]}& 4400.0-4600.0&  $-$1.045 &0.031  \\
& 4600.0-4800.0& $-$0.965 &0.051 \\
& 4800.0-5000.0& $-$0.896 &0.029 \\
& 5000.0-5200.0&  $-$0.955 &0.041 \\
& 6100.0-6300.0& $-$0.985 &0.060\\
& 6300.0-6500.0&  $-$0.934 &0.062 \\
& 6500.0-6700.0&  $-$0.939 &0.093 \\
& 6700.0-6800.0&  $-$1.255 &0.161 \\
&7400.0-7550.0 & $-$0.926 &0.074 \\
&8500.0-8700.0 & $-$0.945 &0.058 \\
& 8700.0-8850.0& $-$0.665 &0.101  \\

{[Mg/Fe]}&4300.0-4370.0& $+$0.531 &0.102 \\
& 4690.0-4710.0 & $+$0.253 &0.177 \\
&5150.0-5200.0& $+$0.616 &0.059\\
&8777.0-8832.0& $+$0.431 &0.073\\

{[Ca/Fe]}&4445.0-4465.0& $+$0.282 &0.128  \\
&6100.0-6128.0& $+$0.562 &0.103\\
&6430.0-6454.0 & $+$0.653 &0.138\\
&6459.0-6478.0 & $+$0.101 &0.233 \\
&8480.0-8586.0& $+$0.292 &0.039 \\
&8623.0-8697.0 & $+$0.333 &0.040 \\

{[Ti/Fe]}&4270.0-4322.0 & $+$0.210 &0.168 \\
& 4440.0-4474.0 &$+$0.152 &0.137  \\
&4650.0-4718.0& $+$0.532 &0.100\\
&4980.0-5045.0& $+$0.282 &0.079 \\
&6584.0-6780.0 & $+$0.708 &0.088\\

{[Na/Fe]}&5670.0-5700.0& - & -\\
& 6148.0-6168.0 & - & -\\

{[Cr/Fe]}&4580.0-4640.0& $+$0.142 &0.117\\
&4640.0-4675.0& $+$0.151 &0.127\\
&4915.0-4930.0& $+$0.852 &0.293 \\

{[Mn/Fe]}& 4450.0-4515.0 & $+$0.241 &0.191 \\
&4750.0-4770.0 & $+$0.123 &0.166 \\

{[Ni/Fe]}& 4700.0-4720.0& $-$0.646 &0.405\\
&4910.0-4955.0&  $+$0.472 &0.128\\
&5075.0-5175.0 & $-$0.138 &0.117\\
&6100.0-6200.0 & $+$0.492 &0.137\\
&6760.0-6800.0 & $+$0.504 &0.217 \\
&7700.0-7800.0 & $+$0.246 &0.257\\
 \hline 
 \end{tabular}
\end{table}

\begin{table}
\caption{Chemical abundances HGHH-23}
\label{tab:8}
 \centering 
\begin{tabular}{cccc} 
 \hline  \hline \
Element& $Wavelength\, [$\AA$]$&Abundance &Error\\
  \hline\
{[Fe/H]}& 4400.0-4600.0&  $-$0.435 &0.011  \\
& 4600.0-4800.0& $-$0.445 &0.030 \\
& 4800.0-5000.0& $-$0.535 &0.010 \\
& 5000.0-5200.0&  $-$0.445 &0.021 \\
& 6100.0-6300.0& $-$0.392 &0.024\\
& 6300.0-6500.0&  $-$0.415 &0.039 \\
& 6500.0-6700.0& $-$0.265 &0.040 \\
& 6700.0-6800.0& $-$0.384 &0.052 \\
&7400.0-7550.0 & $-$0.587 &0.046 \\
&8500.0-8700.0 & $-$0.135 &0.011 \\
& 8700.0-8850.0&  $-$0.313 &0.052  \\

{[Mg/Fe]}&4300.0-4370.0& $+$0.521 &0.051 \\
& 4690.0-4710.0 & $-$0.118 &0.097 \\
&5150.0-5200.0& $+$0.292 &0.042\\
&8777.0-8832.0& $+$0.126 &0.047\\

{[Ca/Fe]}&4445.0-4465.0&  $-$0.117 &0.070  \\
&6100.0-6128.0& $+$0.178 &0.066\\
&6430.0-6454.0 & $+$0.321 &0.077\\
&6459.0-6478.0 & $-$0.426 &0.172 \\
&8480.0-8586.0& $+$0.012 &0.034 \\
&8623.0-8697.0 & $-$0.037 &0.035 \\

{[Ti/Fe]}&4270.0-4322.0 & - & -\\
& 4440.0-4474.0 & $-$0.136 &0.088 \\
&4650.0-4718.0& $+$0.163 &0.062\\
&4980.0-5045.0& $+$0.193 &0.062 \\
&6584.0-6780.0 & $-$0.157 &0.117\\

{[Na/Fe]}&5670.0-5700.0& $+$0.324 &0.127 \\
& 6148.0-6168.0 & $+$0.489 &0.180 \\

{[Cr/Fe]}&4580.0-4640.0& $-$0.287 &0.074\\
&4640.0-4675.0& $+$0.293 &0.058\\
&4915.0-4930.0& $-$0.208 &0.243 \\

{[Mn/Fe]}& 4450.0-4515.0 &$-$0.286 &0.117 \\
&4750.0-4770.0 & $+$0.282 &0.075 \\

{[Ni/Fe]}& 4700.0-4720.0& $+$0.422 &0.112\\
&4910.0-4955.0&  $+$0.220 &0.088\\
&5075.0-5175.0 & $-$0.345 &0.071\\
&6100.0-6200.0 & $-$0.368 &0.103\\
&6760.0-6800.0 & - & -\\
&7700.0-7800.0 & $-$0.458 &0.163\\
 \hline 
 \end{tabular}
\end{table}

\begin{table}
\caption{Chemical abundances HGHH-29}
\label{tab:9}
 \centering 
\begin{tabular}{cccc} 
 \hline  \hline \
Element& $Wavelength\, [$\AA$]$&Abundance &Error\\
  \hline\
{[Fe/H]}& 4400.0-4600.0&  -0.464 &0.029  \\
& 4600.0-4800.0& $-$0.464 &0.032 \\
& 4800.0-5000.0& $-$0.546 &0.025 \\
& 5000.0-5200.0&  $-$0.564 &0.032 \\
& 6100.0-6300.0& $-$0.575 &0.040\\
& 6300.0-6500.0&  $-$0.545 &0.048 \\
& 6500.0-6700.0&  $-$0.474 &0.062 \\
& 6700.0-6800.0& $-$0.634 &0.111 \\
&7400.0-7550.0 & $-$0.555 &0.060 \\
&8500.0-8700.0 & $-$0.405 &0.051 \\
& 8700.0-8850.0&  $-$0.335 &0.061  \\

{[Mg/Fe]}&4300.0-4370.0& $+$0.393 &0.119 \\
& 4690.0-4710.0 & $+$0.194 &0.133 \\
&5150.0-5200.0& $+$0.463 &0.046\\
&8777.0-8832.0& $+$0.425 &0.055\\

{[Ca/Fe]}&4445.0-4465.0&  $-$0.074 0& 0.146  \\
&6100.0-6128.0& $+$0.174 &0.084\\
&6430.0-6454.0 & $+$0.291 &0.110\\
&6459.0-6478.0 & $+$0.150 &0.177 \\
&8480.0-8586.0& $+$0.065 &0.027 \\
&8623.0-8697.0 & $-$0.026 &0.020 \\

{[Ti/Fe]}&4270.0-4322.0 & $+$0.004 &0.151 \\
& 4440.0-4474.0 & $+$0.174 &0.129 \\
&4650.0-4718.0& $+$0.405 &0.104\\
&4980.0-5045.0& $+$0.313 &0.084 \\
&6584.0-6780.0 & $+$0.323 &0.140\\

{[Na/Fe]}&5670.0-5700.0& - & -\\
& 6148.0-6168.0 & - & -\\

{[Cr/Fe]}&4580.0-4640.0& $+$0.075 &0.104\\
&4640.0-4675.0& $+$0.435 &0.095\\
&4915.0-4930.0& $-$0.546 &0.491 \\

{[Mn/Fe]}& 4450.0-4515.0 & $-$0.105 &0.202 \\
&4750.0-4770.0 & $+$0.344 &0.141 \\

{[Ni/Fe]}& 4700.0-4720.0& $+$0.242 &0.189\\
&4910.0-4955.0&  $+$0.423 &0.141\\
&5075.0-5175.0 &  $-$0.116 &0.112\\
&6100.0-6200.0 & $-$0.198 &0.153\\
&6760.0-6800.0 &$-$0.054 &0.203 \\
&7700.0-7800.0 & $+$0.683 &0.171\\
 \hline 
 \end{tabular}
\end{table}

\begin{table}
\caption{Chemical abundances HGHH-34}
\label{tab:10}
 \centering 
\begin{tabular}{cccc} 
 \hline  \hline \
Element& $Wavelength\, [$\AA$]$&Abundance &Error\\
  \hline\
{[Fe/H]}& 4400.0-4600.0&  $-$0.395 &0.021  \\
& 4600.0-4800.0& $-$0.475 &0.043 \\
& 4800.0-5000.0& $-$0.435 &0.011 \\
& 5000.0-5200.0&  $-$0.315 &0.03 \\
& 6100.0-6300.0& $-$0.384 &0.021\\
& 6300.0-6500.0& $-$0.355 &0.047 \\
& 6500.0-6700.0&  $-$0.375 & 0.081 \\
& 6700.0-6800.0&  $-$0.394 &0.082 \\
&7400.0-7550.0 & $-$0.345 &0.050 \\
&8500.0-8700.0 & $-$0.104 &0.042 \\
& 8700.0-8850.0&  $-$0.292 &0.082  \\

{[Mg/Fe]}&4300.0-4370.0& $+$0.381 &0.083 \\
& 4690.0-4710.0 & $+$0.102 &0.122 \\
&5150.0-5200.0& $+$0.482 &0.057\\
&8777.0-8832.0& $+$0.177 &0.069\\

{[Ca/Fe]}&4445.0-4465.0& $+$0.203 &0.105  \\
&6100.0-6128.0& $+$0.263 &0.092\\
&6430.0-6454.0 & $+$0.232 &0.124\\
&6459.0-6478.0 & $+$0.153 &0.193 \\
&8480.0-8586.0& $+$0.002 &0.029 \\
&8623.0-8697.0 & $-$0.047 &0.031 \\

{[Ti/Fe]}&4270.0-4322.0 & $+$0.293 &0.114 \\
& 4440.0-4474.0 & $-$0.158 &0.123 \\
&4650.0-4718.0& $+$0.242 &0.102\\
&4980.0-5045.0& $+$0.942 &0.067 \\
&6584.0-6780.0 & $+$0.183 &0.096\\

{[Na/Fe]}&5670.0-5700.0& - & -\\
& 6148.0-6168.0 & - & -\\

{[Cr/Fe]}&4580.0-4640.0& $-$0.060 &0.102\\
&4640.0-4675.0& $+$0.038 &0.106\\
&4915.0-4930.0& $+$0.503 &0.263 \\

{[Mn/Fe]}& 4450.0-4515.0 & $+$0.112 &0.183 \\
&4750.0-4770.0 & $+$0.221 &0.152 \\

{[Ni/Fe]}& 4700.0-4720.0& $+$0.091 &0.182\\
&4910.0-4955.0&  $+$0.432 &0.115\\
&5075.0-5175.0 &  $+$0.153 &0.085\\
&6100.0-6200.0 & $-$0.058 &0.135\\
&6760.0-6800.0 & $+$0.520 &0.187 \\
&7700.0-7800.0 & $+$0.412 &0.194\\
 \hline 
 \end{tabular}
\end{table}

\begin{table}
\caption{Chemical abundances HGHH-40}
\label{tab:11}
 \centering 
\begin{tabular}{cccc} 
 \hline  \hline \
Element& $Wavelength\, [$\AA$]$&Abundance &Error\\
  \hline\
{[Fe/H]}& 4400.0-4600.0&  $-$1.206 &0.069  \\
& 4600.0-4800.0& $-$1.223 &0.083 \\
& 4800.0-5000.0& $-$1.045 &0.051\\
& 5000.0-5200.0&  $-$1.255 &0.080 \\
& 6100.0-6300.0& $-$1.124 &0.081\\
& 6300.0-6500.0&  $-$1.205 &0.120 \\
& 6500.0-6700.0&  $-$0.928 &0.137 \\
& 6700.0-6800.0&  $-$1.254 &0.233 \\
&7400.0-7550.0 & $-$0.876 &0.127 \\
&8500.0-8700.0 & $-$1.155 &0.104 \\
& 8700.0-8850.0&  $-$1.238 &0.168  \\

{[Mg/Fe]}&4300.0-4370.0& $+$0.118 &0.332 \\
& 4690.0-4710.0 & $+$0.108 &0.362 \\
&5150.0-5200.0& $+$0.337 &0.097\\
&8777.0-8832.0& $-$0.073 &0.174\\

{[Ca/Fe]}&4445.0-4465.0&  $-$0.162 &0.374  \\
&6100.0-6128.0& $+$0.286 &0.251\\
&6430.0-6454.0 & $+$0.116 &0.302\\
&6459.0-6478.0 & $+$0.197 &0.333 \\
&8480.0-8586.0& $+$0.197 &0.046 \\
&8623.0-8697.0 & $+$0.177 &0.058 \\

{[Ti/Fe]}&4270.0-4322.0 & $-$0.002 &0.254 \\
& 4440.0-4474.0 & $-$0.313 &0.293 \\
&4650.0-4718.0& $+$0.118 &0.222\\
&4980.0-5045.0& $+$0.478 &0.124 \\
&6584.0-6780.0 & - & - \\

{[Na/Fe]}&5670.0-5700.0& - & -\\
& 6148.0-6168.0 & - & -\\

{[Cr/Fe]}&4580.0-4640.0& $+$0.207 &0.193\\
&4640.0-4675.0& $-$0.593 &0.382\\
&4915.0-4930.0& - & - \\

{[Mn/Fe]}& 4450.0-4515.0 & $-$0.072 &0.393 \\
&4750.0-4770.0 & $-$0.173 &0.308 \\

{[Ni/Fe]}& 4700.0-4720.0& - & - \\
&4910.0-4955.0&  $+$0.688 &0.190\\
&5075.0-5175.0 & $-$1.023 &0.472\\
&6100.0-6200.0 & $+$0.068 &0.254\\
&6760.0-6800.0 & $+$0.206 &0.350 \\
&7700.0-7800.0 & $+$0.246 &0.242\\
 \hline 
 \end{tabular}
\end{table}

\begin{table}
\caption{Chemical abundances HH-080}
\label{tab:12}
 \centering 
\begin{tabular}{cccc} 
 \hline  \hline \
Element& $Wavelength\, [$\AA$]$&Abundance &Error\\
  \hline\
{[Fe/H]}& 4400.0-4600.0&  $-$1.825 &0.058  \\
& 4600.0-4800.0& $-$1.716 &0.075 \\
& 4800.0-5000.0& $-$1.696 &0.036 \\
& 5000.0-5200.0&  $-$1.696 &0.045 \\
& 6100.0-6300.0& $-$1.536 &0.088\\
& 6300.0-6500.0& $-$1.286 &0.077 \\
& 6500.0-6700.0& $-$1.454 &0.122 \\
& 6700.0-6800.0& $-$1.326 &0.442 \\
&7400.0-7550.0 & $-$1.192 &0.096 \\
&8500.0-8700.0 & $-$1.714 &0.102 \\
& 8700.0-8850.0& $-$1.465 &0.161  \\

{[Mg/Fe]}&4300.0-4370.0& $+$0.399&0.239  \\
& 4690.0-4710.0 & $-$0.293 &0.256 \\
&5150.0-5200.0& $-$0.162 &0.100\\
&8777.0-8832.0& $+$0.357 &0.159\\

{[Ca/Fe]}&4445.0-4465.0& $+$0.057 &0.198  \\
&6100.0-6128.0& $-$0.110 &0.211\\
&6430.0-6454.0 & $+$0.0157 &0.227\\
&6459.0-6478.0 & $+$0.335 &0.255 \\
&8480.0-8586.0&  $+$0.298 &0.061 \\
&8623.0-8697.0 & $+$0.317 &0.061 \\

{[Ti/Fe]}&4270.0-4322.0 & $-$0.083 &0.170 \\
& 4440.0-4474.0 & $-$0.223 &0.199 \\
&4650.0-4718.0& $+$0.077 &0.180\\
&4980.0-5045.0& $+$0.199 &0.094 \\
&6584.0-6780.0 &  $+$0.588 &0.322\\

{[Na/Fe]}&5670.0-5700.0& - & -\\
& 6148.0-6168.0 & - & -\\

{[Cr/Fe]}&4580.0-4640.0& $-$0.003 &0.180\\
&4640.0-4675.0& $-$0.154 &0.246\\
&4915.0-4930.0& $-$0.081 &0.500 \\

{[Mn/Fe]}& 4450.0-4515.0 & $+$0.188 &0.263 \\
&4750.0-4770.0 &  $-$0.184 &0.198 \\

{[Ni/Fe]}& 4700.0-4720.0& $-$0.242 &0.390\\
&4910.0-4955.0&  $+$0.096 &0.380\\
&5075.0-5175.0 &  $-$0.043 &0.149\\
&6100.0-6200.0 &  $+$0.587 &0.204\\
&6760.0-6800.0 & $+$0.010 &0.329 \\
&7700.0-7800.0 & $+$0.204 &0.312\\
 \hline 
 \end{tabular}
\end{table}

\begin{table}
\caption{Chemical abundances HH-096}
\label{tab:13}
 \centering 
\begin{tabular}{cccc} 
 \hline  \hline \
Element& $Wavelength\, [$\AA$]$&Abundance &Error\\
  \hline\
{[Fe/H]}& 4400.0-4600.0&  $-$1.265 &0.041  \\
& 4600.0-4800.0& $-$1.185 &0.041 \\
& 4800.0-5000.0& $-$1.246 &0.039 \\
& 5000.0-5200.0& $-$1.424 &0.042 \\
& 6100.0-6300.0& $-$1.096 &0.058\\
& 6300.0-6500.0&  $-$1.295 &0.070 \\
& 6500.0-6700.0&  $-$1.496 &0.097 \\
& 6700.0-6800.0& $-$1.194 &0.162 \\
&7400.0-7550.0 & $-$1.547 &0.109 \\
&8500.0-8700.0 & $-$1.548 &0.067 \\
& 8700.0-8850.0& $-$1.177 &0.118  \\

{[Mg/Fe]}&4300.0-4370.0& $+$0.704 &0.108 \\
& 4690.0-4710.0 & $+$0.317 &0.217 \\
&5150.0-5200.0& $+$0.337 &0.068\\
&8777.0-8832.0& $-$0.323 &0.157\\

{[Ca/Fe]}&4445.0-4465.0&  $+$0.216 &0.167  \\
&6100.0-6128.0& $+$0.375 &0.147\\
&6430.0-6454.0 & $+$0.398 &0.169\\
&6459.0-6478.0 & $+$0.287 &0.209 \\
&8480.0-8586.0& $+$0.406 &0.046 \\
&8623.0-8697.0 & $+$0.287 &0.049 \\

{[Ti/Fe]}&4270.0-4322.0 & $+$0.236 &0.147 \\
& 4440.0-4474.0 & $+$0.407 &0.139 \\
&4650.0-4718.0& $+$0.347 &0.120\\
&4980.0-5045.0& $+$0.166 &0.092 \\
&6584.0-6780.0 & $+$0.454 &0.229\\

{[Na/Fe]}&5670.0-5700.0& - & -\\
& 6148.0-6168.0 & - & -\\

{[Cr/Fe]}&4580.0-4640.0& $+$0.086 &0.127\\
&4640.0-4675.0& $+$0.025 &0.185\\
&4915.0-4930.0& $+$0.377 &0.394 \\

{[Mn/Fe]}& 4450.0-4515.0 & $-$0.624 &0.502\\
&4750.0-4770.0 & $-$0.194 &0.186 \\

{[Ni/Fe]}& 4700.0-4720.0& $-$0.213 &0.326\\
&4910.0-4955.0& $-$0.141 &0.262\\
&5075.0-5175.0 & $+$0.195 &0.120\\
&6100.0-6200.0 & $+$0.126 &0.206\\
&6760.0-6800.0 & $+$0.403 &0.232 \\
&7700.0-7800.0 & $+$0.016 &0.249\\
 \hline 
 \end{tabular}
\end{table}

\begin{table}
\caption{Chemical abundances HHH86-30}
\label{tab:14}
 \centering 
\begin{tabular}{cccc} 
 \hline  \hline \
Element& $Wavelength\, [$\AA$]$&Abundance &Error\\
  \hline\
{[Fe/H]}& 4400.0-4600.0&  $-$0.316 &0.029  \\
& 4600.0-4800.0& $-$0.275 &0.011 \\
& 4800.0-5000.0& $-$0.355 &0.021 \\
& 5000.0-5200.0& $-$0.335 &0.011 \\
& 6100.0-6300.0& $-$0.187 &0.038\\
& 6300.0-6500.0& $-$0.396 &0.030 \\
& 6500.0-6700.0& $-$0.275 &0.041 \\
& 6700.0-6800.0& $-$0.564 &0.062 \\
&7400.0-7550.0 & $-$0.334 &0.032 \\
&8500.0-8700.0 & $+$0.035 &0.021 \\
& 8700.0-8850.0& $-$0.395 &0.051  \\

{[Mg/Fe]}&4300.0-4370.0& $+$0.240 &0.058 \\
& 4690.0-4710.0 & $+$0.129 &0.073 \\
&5150.0-5200.0& $+$0.371 &0.046\\
&8777.0-8832.0& $+$0.231 &0.047\\

{[Ca/Fe]}&4445.0-4465.0& $-$0.009 &0.082  \\
&6100.0-6128.0& $+$0.340 &0.073\\
&6430.0-6454.0 & $+$0.399 &0.081\\
&6459.0-6478.0 &$-$0.308 &0.168 \\
&8480.0-8586.0& $-$0.129 &0.042 \\
&8623.0-8697.0 & $-$0.190 &0.041 \\

{[Ti/Fe]}&4270.0-4322.0 & $+$0.062 &0.089 \\
& 4440.0-4474.0 & $-$0.170 &0.090 \\
&4650.0-4718.0& $+$0.281 &0.066\\
&4980.0-5045.0& $+$0.421 &0.059 \\
&6584.0-6780.0 & $+$0.372 &0.092\\

{[Na/Fe]}&5670.0-5700.0& $+$0.199 &0.138 \\
& 6148.0-6168.0 & $+$0.390 &0.203 \\

{[Cr/Fe]}&4580.0-4640.0& $-$0.039 &0.066\\
&4640.0-4675.0& $+$0.425 &0.059\\
&4915.0-4930.0& $+$0.460 &0.155 \\

{[Mn/Fe]}& 4450.0-4515.0 & $+$0.268 &0.120 \\
&4750.0-4770.0 & $+$0.130 &0.099 \\

{[Ni/Fe]}& 4700.0-4720.0& $+$0.061 &0.109\\
&4910.0-4955.0& $+$0.071 &0.082\\
&5075.0-5175.0 & $+$0.091 &0.058\\
&6100.0-6200.0 & $+$0.181 &0.090\\
&6760.0-6800.0 & $+$0.232 &0.138 \\
&7700.0-7800.0 & $-$0.130 &0.165\\
 \hline 
 \end{tabular}
\end{table}

\begin{table}
\caption{Chemical abundances HHH86-39}
\label{tab:15}
 \centering 
\begin{tabular}{cccc} 
 \hline  \hline \
Element& $Wavelength\, [$\AA$]$&Abundance &Error\\
  \hline\
{[Fe/H]}& 4400.0-4600.0&  $-$1.064 &0.012  \\
& 4600.0-4800.0& $-$1.554 &0.031 \\
& 4800.0-5000.0& $-$1.024 &0.021 \\
& 5000.0-5200.0&  $-$1.174 &0.021 \\
& 6100.0-6300.0&  $-$1.525 &0.021\\
& 6300.0-6500.0&  $-$1.495 &0.030 \\
& 6500.0-6700.0&  $-$1.574 &0.062 \\
& 6700.0-6800.0&  $-$1.913 &0.146 \\
&7400.0-7550.0 & $-$1.354 &0.031 \\
&8500.0-8700.0 & $-$1.595 & 0.031 \\
& 8700.0-8850.0&  $-$1.625 &0.041  \\

{[Mg/Fe]}&4300.0-4370.0& -  & -\\
& 4690.0-4710.0 &-  & - \\
&5150.0-5200.0&-  & -\\
&8777.0-8832.0&-  & -\\

{[Ca/Fe]}&4445.0-4465.0&  $+$1.117 &0.090  \\
&6100.0-6128.0& $+$0.727 &0.092\\
&6430.0-6454.0 & $+$0.425 &0.110\\
&6459.0-6478.0 & $+$0.013 &0.124 \\
&8480.0-8586.0& $+$0.077 &0.074 \\
&8623.0-8697.0 & $+$0.028 &0.076 \\

{[Ti/Fe]}&4270.0-4322.0 & $+$0.437 &0.089 \\
& 4440.0-4474.0 & $-$0.004&0.120 \\
&4650.0-4718.0& $-$0.650 &0.144\\
&4980.0-5045.0& $+$0.266 &0.084 \\
&6584.0-6780.0 & $-$0.702 &0.282\\

{[Na/Fe]}&5670.0-5700.0& $-$0.032 &0.125 \\
& 6148.0-6168.0 & $-$0.191 &0.282 \\

{[Cr/Fe]}&4580.0-4640.0& $-$0.371 &0.110\\
&4640.0-4675.0& $-$0.222 &0.118\\
&4915.0-4930.0& - & -\\

{[Mn/Fe]}& 4450.0-4515.0 & $-$0.759 &0.248 \\
&4750.0-4770.0 & $-$0.252 &0.119 \\

{[Ni/Fe]}& 4700.0-4720.0& $+$0.357 &0.117\\
&4910.0-4955.0&  $-$0.043 &0.128\\
&5075.0-5175.0 & $-$0.153 &0.095\\
&6100.0-6200.0 & $-$0.442&0.148\\
&6760.0-6800.0 & $-$0.335 &0.162 \\
&7700.0-7800.0 & $-$0.565 &0.140\\
 \hline 
 \end{tabular}
\end{table}

\begin{table}
\caption{Chemical abundances K-029}
\label{tab:16}
 \centering 
\begin{tabular}{cccc} 
 \hline  \hline \
Element& $Wavelength\, [$\AA$]$&Abundance &Error\\
  \hline\
{[Fe/H]}& 4400.0-4600.0&  $-$0.774 &0.021  \\
& 4600.0-4800.0& $-$0.805 &0.041 \\
& 4800.0-5000.0& $-$0.775 &0.030 \\
& 5000.0-5200.0&  $-$0.744 &0.021 \\
& 6100.0-6300.0& $-$0.696 &0.046\\
& 6300.0-6500.0&  $-$0.690 &0.050 \\
& 6500.0-6700.0&  $-$0.714 &0.062 \\
& 6700.0-6800.0&  $-$0.644 &0.111  \\
&7400.0-7550.0 & $-$0.685 &0.062 \\
&8500.0-8700.0 & $-$0.664 &0.042 \\
& 8700.0-8850.0&  $-$0.507 &0.078  \\

{[Mg/Fe]}&4300.0-4370.0& $+$0.767 &0.074 \\
& 4690.0-4710.0 & $+$0.358 &0.114 \\
&5150.0-5200.0& $+$0.406 &0.046\\
&8777.0-8832.0& $+$0.317 &0.061\\

{[Ca/Fe]}&4445.0-4465.0&  $-$0.0435 &0.141  \\
&6100.0-6128.0& $+$0.513 &0.080\\
&6430.0-6454.0 & $+$0.360 &0.118\\
&6459.0-6478.0 & $-$0.437 &0.248 \\
&8480.0-8586.0& $+$0.168 &0.025 \\
&8623.0-8697.0 & $+$0.138 &0.025 \\

{[Ti/Fe]}&4270.0-4322.0 & $-$0.123 &0.146 \\
& 4440.0-4474.0 & $-$0.245 &0.148 \\
&4650.0-4718.0& $+$0.328 &0.094\\
&4980.0-5045.0& $+$0.328 &0.065 \\
&6584.0-6780.0 & $+$0.177 &0.096\\

{[Na/Fe]}&5670.0-5700.0& $+$0.636 &0.176 \\
& 6148.0-6168.0 & $+$0.797 &0.216 \\

{[Cr/Fe]}&4580.0-4640.0& $-$0.153 &0.095\\
&4640.0-4675.0& $+$0.298 &0.084\\
&4915.0-4930.0& $+$0.707 &0.212 \\

{[Mn/Fe]}& 4450.0-4515.0 & $-$0.662 &0.243 \\
&4750.0-4770.0 & $+$0.327 &0.131 \\

{[Ni/Fe]}& 4700.0-4720.0& $+$0.238 &0.144\\
&4910.0-4955.0& $+$0.258 &0.113\\
&5075.0-5175.0 & $-$0.3834 &0.103\\
&6100.0-6200.0 & $+$0.150 &0.124\\
&6760.0-6800.0 & $+$0.007 &0.212 \\
&7700.0-7800.0 & $-$0.284 &0.245\\
 \hline 
 \end{tabular}
\end{table}

\begin{table}
\caption{Chemical abundances K-034}
\label{tab:17}
 \centering 
\begin{tabular}{cccc} 
 \hline  \hline \
Element& $Wavelength\, [$\AA$]$&Abundance &Error\\
  \hline\
{[Fe/H]}& 4400.0-4600.0&  $-$0.534 &0.012  \\
& 4600.0-4800.0& $-$0.485 &0.030 \\
& 4800.0-5000.0& $-$0.596 &0.024 \\
& 5000.0-5200.0&  $-$0.534 &0.012 \\
& 6100.0-6300.0& $-$0.654 &0.031\\
& 6300.0-6500.0&  $-$0.594 &0.035 \\
& 6500.0-6700.0& $-$0.445 &0.049 \\
& 6700.0-6800.0&  $-$0.604 &0.081 \\
&7400.0-7550.0 & $-$0.535 &0.042 \\
&8500.0-8700.0 & $-$0.325 &0.021 \\
& 8700.0-8850.0& $-$0.305 &0.054  \\

{[Mg/Fe]}&4300.0-4370.0& $+$0.723 &0.049 \\
& 4690.0-4710.0 & $+$0.213 &0.087 \\
&5150.0-5200.0& $+$0.391 &0.049\\
&8777.0-8832.0& $+$0.263 &0.044\\

{[Ca/Fe]}&4445.0-4465.0& $+$0.014 &0.097  \\
&6100.0-6128.0& $+$0.164 &0.077\\
&6430.0-6454.0 & $+$0.471 &0.090\\
&6459.0-6478.0 & $+$0.064 &0.152 \\
&8480.0-8586.0& $-$0.007 &0.031 \\
&8623.0-8697.0 & $-$0.027 &0.032 \\

{[Ti/Fe]}&4270.0-4322.0 & $+$0.113 &0.090 \\
& 4440.0-4474.0 & $+$0.045 &0.092 \\
&4650.0-4718.0& $+$0.104 &0.097\\
&4980.0-5045.0& $+$0.304 &0.052 \\
&6584.0-6780.0 & $-$0.002 &0.168\\

{[Na/Fe]}&5670.0-5700.0& $+$0.293 &0.144 \\
& 6148.0-6168.0 & $+$0.432 &0.192 \\

{[Cr/Fe]}&4580.0-4640.0& $-$0.017 &0.068\\
&4640.0-4675.0& $+$0.403 &0.060\\
&4915.0-4930.0& $+$0.614&0.185 \\

{[Mn/Fe]}& 4450.0-4515.0 & $+$0.144 &0.135 \\
&4750.0-4770.0 & $+$0.142 &0.109 \\

{[Ni/Fe]}& 4700.0-4720.0& $+$0.233 &0.124\\
&4910.0-4955.0&  $+$0.447 &0.087\\
&5075.0-5175.0 &  $-$0.227 &0.085\\
&6100.0-6200.0 & $-$0.087 &0.112\\
&6760.0-6800.0 & $+$0.267 &0.154 \\
&7700.0-7800.0 & $-$0.107 &0.202\\
 \hline 
 \end{tabular}
\end{table}

\begin{table}
\caption{Chemical abundances K-163}
\label{tab:18}
 \centering 
\begin{tabular}{cccc} 
 \hline  \hline \
Element& $Wavelength\, [$\AA$]$&Abundance &Error\\
  \hline\
{[Fe/H]}& 4400.0-4600.0&  $-$1.115 &0.012  \\
& 4600.0-4800.0& $-$1.049 &0.036 \\
& 4800.0-5000.0& $-$1.145 &0.021 \\
& 5000.0-5200.0&  $-$1.095 &0.021 \\
& 6100.0-6300.0& $-$1.060 &0.046\\
& 6300.0-6500.0&  $-$1.026 &0.043 \\
& 6500.0-6700.0&  $-$0.805 &0.057 \\
& 6700.0-6800.0&$-$1.115 &0.157 \\
&7400.0-7550.0 & $-$1.196 &0.210 \\
&8500.0-8700.0 & $-$1.136 &0.064 \\
& 8700.0-8850.0&  $-$0.876 &0.079  \\

{[Mg/Fe]}&4300.0-4370.0& $+$0.867 &0.051 \\
& 4690.0-4710.0 & $+$0.289 &0.128 \\
&5150.0-5200.0& $+$0.448 &0.048\\
&8777.0-8832.0& $+$0.527 &0.057\\

{[Ca/Fe]}&4445.0-4465.0& $+$0.157 &0.101  \\
&6100.0-6128.0&  $+$0.527 &0.089\\
&6430.0-6454.0 & $+$0.753 &0.101\\
&6459.0-6478.0 & $+$0.440 &0.158 \\
&8480.0-8586.0& $+$0.458 &0.037 \\
&8623.0-8697.0 & $+$0.401&0.037 \\

{[Ti/Fe]}&4270.0-4322.0 & $+$0.077 &0.107 \\
& 4440.0-4474.0 & $+$0.146 &0.101 \\
&4650.0-4718.0& $+$0.338 &0.079\\
&4980.0-5045.0& $+$0.328 &0.060 \\
&6584.0-6780.0 & $+$0.254 &0.096\\

{[Na/Fe]}&5670.0-5700.0& - & -\\
& 6148.0-6168.0 & $+$0.648 &0.219 \\

{[Cr/Fe]}&4580.0-4640.0& $-$0.083 &0.089\\
&4640.0-4675.0& $+$0.097 &0.107\\
&4915.0-4930.0& $+$0.547 &0.253 \\

{[Mn/Fe]}& 4450.0-4515.0 & $-$0.453 &0.184 \\
&4750.0-4770.0 & $-$0.113 &0.116 \\

{[Ni/Fe]}& 4700.0-4720.0& $+$0.126 &0.164\\
&4910.0-4955.0& $+$0.146 &0.120\\
&5075.0-5175.0 &  $-$0.223 &0.079\\
&6100.0-6200.0 & $+$0.438 &0.111\\
&6760.0-6800.0 & $+$0.387 &0.191 \\
&7700.0-7800.0 & $+$0.637 &0.179\\
 \hline 
 \end{tabular}
\end{table}

\begin{table}
\caption{Chemical abundances VHH81-03}
\label{tab:19}
 \centering 
\begin{tabular}{cccc} 
 \hline  \hline \
Element& $Wavelength\, [$\AA$]$&Abundance &Error\\
  \hline\
{[Fe/H]}& 4400.0-4600.0&  $-$0.234 &0.012  \\
& 4600.0-4800.0& $-$0.135 &0.021 \\
& 4800.0-5000.0& $-$0.166 &0.021 \\
& 5000.0-5200.0&  $-$0.165 &0.021 \\
& 6100.0-6300.0& $-$0.175 &0.035\\
& 6300.0-6500.0&  $-$0.205 &0.040 \\
& 6500.0-6700.0&  $-$0.344 &0.071 \\
& 6700.0-6800.0&  $-$0.094 &0.062 \\
&7400.0-7550.0 & $-$0.263 &0.042 \\
&8500.0-8700.0 & $+$0.005 &0.011 \\
& 8700.0-8850.0&  $-$0.045 &0.053  \\

{[Mg/Fe]}&4300.0-4370.0& $+$0.343 &0.075 \\
& 4690.0-4710.0 & $+$0.188 &0.085 \\
&5150.0-5200.0& $+$0.392 &0.058\\
&8777.0-8832.0& $+$0.106 &0.051\\

{[Ca/Fe]}&4445.0-4465.0& $+$0.203&0.091  \\
&6100.0-6128.0& $+$0.359 &0.079\\
&6430.0-6454.0 & $+$0.410&0.098\\
&6459.0-6478.0 & $-$0.505 &0.207 \\
&8480.0-8586.0& $-$0.156 &0.045 \\
&8623.0-8697.0 & $-$0.217 &0.044 \\

{[Ti/Fe]}&4270.0-4322.0 & $-$0.037 &0.091 \\
& 4440.0-4474.0 & $+$0.064 &0.093 \\
&4650.0-4718.0& $+$0.103 &0.075\\
&4980.0-5045.0& $+$0.213&0.059 \\
&6584.0-6780.0 & $-$0.309 &0.167\\

{[Na/Fe]}&5670.0-5700.0& - & -\\
& 6148.0-6168.0 & - & -\\

{[Cr/Fe]}&4580.0-4640.0& $-$0.126 &0.075\\
&4640.0-4675.0& $+$0.286 &0.076\\
&4915.0-4930.0& $+$0.874 &0.157 \\

{[Mn/Fe]}& 4450.0-4515.0 & $-$0.182 &0.144 \\
&4750.0-4770.0 & $+$0.393 &0.104 \\

{[Ni/Fe]}& 4700.0-4720.0& $-$0.277 &0.156\\
&4910.0-4955.0&  $+$0.075 &0.100\\
&5075.0-5175.0 &  $+$0.132 &0.081\\
&6100.0-6200.0 & $+$0.033 &0.110\\
&6760.0-6800.0 & $+$0.232 &0.175 \\
&7700.0-7800.0 & $+$0.398 &0.161\\
 \hline 
 \end{tabular}
\end{table}

\begin{table}
\caption{Chemical abundances VHH81-05}
\label{tab:20}
 \centering 
\begin{tabular}{cccc} 
 \hline  \hline \
Element& $Wavelength\, [$\AA$]$&Abundance &Error\\
  \hline\
{[Fe/H]}& 4400.0-4600.0&  $-$1.643 &0.053  \\
& 4600.0-4800.0& $-$1.683 &0.072 \\
& 4800.0-5000.0& $-$1.564 &0.042 \\
& 5000.0-5200.0&  $-$1.465 &0.051 \\
& 6100.0-6300.0& $-$1.403 &0.083\\
& 6300.0-6500.0&  $-$1.623 &0.092 \\
& 6500.0-6700.0& $-$2.074 &0.111 \\
& 6700.0-6800.0&  $-$1.653 &0.313 \\
&7400.0-7550.0 &$-$1.413 &0.123 \\
&8500.0-8700.0 & $-$1.944 &0.071 \\
& 8700.0-8850.0&  $-$1.423 &0.163  \\

{[Mg/Fe]}&4300.0-4370.0& $+$0.396 & 0.110 \\
& 4690.0-4710.0 & $+$0.156 &0.278 \\
&5150.0-5200.0& $+$0.246 &0.079\\
&8777.0-8832.0& $-$0.074 &0.162\\

{[Ca/Fe]}&4445.0-4465.0& $+$0.227&0.211  \\
&6100.0-6128.0& $+$0.487 &0.182\\
&6430.0-6454.0 & $+$0.266&0.239\\
&6459.0-6478.0 & - & -  \\
&8480.0-8586.0& $+$0.546 &0.061 \\
&8623.0-8697.0 & $+$0.526 &0.063 \\

{[Ti/Fe]}&4270.0-4322.0 & $+$0.356&0.162 \\
& 4440.0-4474.0 & $+$0.456 &0.154 \\
&4650.0-4718.0& $+$0.277 &0.163\\
&4980.0-5045.0& $+$0.507 &0.110 \\
&6584.0-6780.0 & $+$1.176 &0.154\\

{[Na/Fe]}&5670.0-5700.0& - & -\\
& 6148.0-6168.0 & - & -\\

{[Cr/Fe]}&4580.0-4640.0& $-$0.134 &0.219\\
&4640.0-4675.0& $-$0.053 &0.299\\
&4915.0-4930.0& - & -\\

{[Mn/Fe]}& 4450.0-4515.0 & $-$0.163 &0.397 \\
&4750.0-4770.0 & $-$0.194 &0.229 \\

{[Ni/Fe]}& 4700.0-4720.0& $-$0.753 &0.725\\
&4910.0-4955.0&  $-$0.454 &0.982\\
&5075.0-5175.0 & $-$0.465 &0.219\\
&6100.0-6200.0 & - & -\\
&6760.0-6800.0 & $-$0.144 &0.346 \\
&7700.0-7800.0 & $-$0.464 &0.804\\
 \hline 
 \end{tabular}
\end{table}

\bsp	
\label{lastpage}
\end{document}